# Diffusion in Lorentz Lattice Gas Cellular Automata: the honeycomb and quasi-lattices compared with the square and triangular lattices

F. Wang
and
E. G. D. Cohen
The Rockefeller University
New York, NY 10021, USA

**Abstract**

We study numerically the nature of the diffusion process on a honeycomb and a quasi-lattice, where a point particle, moving along the bonds of the lattice, scatters from randomly placed scatterers on the lattice sites according to strictly deterministic rules. For the honeycomb lattice fully occupied by fixed rotators two (symmetric) isolated critical points appear to be present, with the same hyperscaling relation as for the square and the triangular lattices. No such points appear to exist for the quasi-lattice. A comprehensive comparison is made with the behavior on the previously studied square and triangular lattices. A great variety of diffusive behavior is found ranging from propagation, super-diffusion, normal, quasi-normal, anomalous to absence of diffusion. The influence of the scattering rules as well as of the lattice structure on the diffusive behavior of a point particle moving on the all lattices studied so far, is summarized.

# 1 Introduction

In a number of previous papers we have studied numerically the diffusion in Lorentz Lattice Gas Cellular Automata (LLGCA) for a variety of strictly deterministic scattering rules[1−7] (for reviews see refs.[1], [4] and [6]). In particular new types of diffusion were found on the square and triangular lattices[1]. In order to see to what extent the nature of the diffusion process depends on the type of lattice, we report here results for the honeycomb and quasi-lattices. Although some preliminary investigations were also made on the random lattice, they will not be considered here since they were not elaborate enough to be used for comparison[2]. The random lattice will be discussed in a later publication[8]. Since not all lattices considered in this paper are regular lattices, the results obtained allow us also a comparison of the diffusive behavior on a lattice in its dependence on the lattice structure as well as on the scattering rules.

In LLGCA a point particle moves along the bonds of a lattice, whose sites are randomly occupied by stationary (in position) scatterers, which scatter the particle according to deterministic rules. We have studied two models for the scatterers, with about the simplest nontrivial scattering rules one can think of: either with right and left mirrors or with right and left rotators and considered two cases for each model: the scatterers do or do not change in character at collision, respectively. In the first case we consider fixed scatterers. i.e., stationary both in position and character (discussed in ref.[1]), in the second case we consider flipping scatterers, i.e., stationary in position, but changing from right to left and vice-versa after a collision.

In the case of fixed scatterers one can consider one particle or many particles moving simultaneously but independently of each other through the scatterers. For flipping scatterers, however, the motion of more than one particle through the scatterers differs qualitatively from that of one single particle, since the flipping of the scatterers introduces (indirect) interactions between the particles as well as the scatterers. We confine ourselves here to the motion of a single particle.

Since straight lines do not pass through any lattice site on the honeycomb lattice and pass through some but not all lattice sites on the quasi-lattice, only the case of a honeycomb lattice and a quasi-lattice fully occupied by scatterers, i.e., with concentration of scatterers $C = 1$, has been considered, since otherwise a mixture of two models with different scattering rules would have to be introduced.

The results of the investigations reported in this paper, which were carried out to over a million or more time steps, can be summarized as follows.

1. Like on the square (for the fixed rotator model) and the triangular (for both the fixed rotator and the fixed mirror models) lattices, there appear to exist critical points for the fixed rotator model on the honeycomb lattice at which the diffusive behavior is anomalous and the trajectories obey a hyperscaling relation, while there is an absence of diffusion–to which we will refer as no-diffusion–for all other concentrations. There is no-diffusion for any concentration of left or right fixed rotators on the quasi-lattice.

2. There is no-diffusion for flipping scatterers (mirrors or rotators) on both the honeycomb and the quasi-lattices.

3. A more detailed summary and comparison of the results for all lattices studied so far is given in Section 4.

The organization of this paper is as follows. In Section 2, we study the honeycomb lattice for the two models mentioned above, by specifying the equations of motion for the moving particle. The Boltzmann approximation to the diffusion coefficient is given, the computer simulation method is described, and the results of the simulations for both fixed and flipping scatterers are discussed. Section 3 discusses how to construct a Fibonacci quasi-lattice as well as the equations of motion for a particle moving on this lattice. The Boltzmann approximation is given, and the computer simulations as well as their results are discussed. In Section 4 we give a summary and a comparison with previous work.

# 2 Honeycomb Lattice

## A. Equations of Motion

The honeycomb lattice, shown in fig.1, has two kinds of sites which, after a rotation over $\pi$, transform into each other (cf.fig.2a) and play an identical role in the diffusion process. The scattering rules for the mirror and the rotator models are shown in fig.2b-c[1]. They lead to the following equations of motion for a particle for both fixed and flipping rotators:

$$\begin{aligned}
n_1(\vec{r}+\vec{e}_1,t+1) &= n_2(\vec{r},t)m_R(\vec{r},t)+n_6(\vec{r},t)m_L(\vec{r},t) \\
n_i(\vec{r}+\vec{e}_i,t+1) &= n_{i+1}(\vec{r},t)m_R(\vec{r},t)+n_{i-1}(\vec{r},t)m_L(\vec{r},t) \qquad (i=2,3,4,5) \\
n_6(\vec{r}+\vec{e}_6,t+1) &= n_1(\vec{r},t)m_R(\vec{r},t)+n_5(\vec{r},t)m_L(\vec{r},t) \qquad (2.1)
\end{aligned}$$

while for both fixed and flipping mirrors, one obtains:

$$\begin{aligned}
n_1(\vec{r}+\vec{e}_1,t+1) &= n_2(\vec{r},t)m_R(\vec{r},t)+n_6(\vec{r},t)m_L(\vec{r},t) \\
n_i(\vec{r}+\vec{e}_i,t+1) &= n_{i-1}(\vec{r},t)m_R(\vec{r},t)+n_{i+1}(\vec{r},t)m_L(\vec{r},t) \qquad (i=2,4) \\
n_i(\vec{r}+\vec{e}_i,t+1) &= n_{i-1}(\vec{r},t)m_L(\vec{r},t)+n_{i+1}(\vec{r},t)m_R(\vec{r},t) \qquad (i=3,5) \\
n_6(\vec{r}+\vec{e}_6,t+1) &= n_1(\vec{r},t)m_L(\vec{r},t)+n_5(\vec{r},t)m_R(\vec{r},t) \qquad (2.2)
\end{aligned}$$

Here, the $\vec{e}_i$ are the unit vectors along the six velocity directions $i=1,...,6$ (see fig.2); $n_i(\vec{r},t)=1$ or $0$ ($i=1,...,6$), if a particle is or is not at the lattice site $\vec{r}$ at time $t$, respectively; $m_{R,L}(\vec{r},t)=1$ or $0$, if a right (left) rotator (in the rotator model) or a right (left) mirror (in the mirror model) is or is not at the lattice site $\vec{r}$ at time $t$, respectively. $m_R(\vec{r},t)$ or $m_L(\vec{r},t)$ are independent of $t$ for fixed scatterers, while they depend on $t$ for flipping scatterers.

[1]We remark that for the fully occupied triangular lattice, if the moving particle turns $\pm\pi/3$ upon collision with a scatterer, the particle moves on the honeycomb lattice considered here[2].

B. Boltzmann Approximation

The Boltzmann approximation to the eqs.(2.1) – (2.2) is obtained by averaging both sides of the equations over all possible random configurations of the scatterer and the particle and ignoring any correlations between scatterer occupation and particle velocity at each lattice site. This leads for all cases, i.e., for fixed and flipping mirrors or rotators to an equation of the form:

$$f_i(t+1) = f_i(t) + \sum_{j=1}^{6} T_{ij} f_j(t) \quad (i = 1, ..., 6) \tag{2.3}$$

where $f_i(t)$ is the probability to find a particle with the velocity direction along $\vec{e}_i$ at time $t$ and $T_{ij}$ are the elements of the collision matrix $\hat{T}$, given by:

$$\hat{T}^{mirror} = \begin{pmatrix} -1 & C_R & 0 & 0 & 0 & C_L \\ C_R & -1 & C_L & 0 & 0 & 0 \\ 0 & C_L & -1 & C_R & 0 & 0 \\ 0 & 0 & C_R & -1 & C_L & 0 \\ 0 & 0 & 0 & C_L & -1 & C_R \\ C_L & 0 & 0 & 0 & C_R & -1 \end{pmatrix}$$

for the mirror case and

$$\hat{T}^{totator} = \begin{pmatrix} -1 & C_R & 0 & 0 & 0 & C_L \\ C_L & -1 & C_R & 0 & 0 & 0 \\ 0 & C_L & -1 & C_R & 0 & 0 \\ 0 & 0 & C_L & -1 & C_R & 0 \\ 0 & 0 & 0 & C_L & -1 & C_R \\ C_R & 0 & 0 & 0 & C_L & -1 \end{pmatrix}$$

for the rotator case, where $C_{L,R} =< m_{L,R}(\vec{r},t) >$, the concentration of left (right) scatterers on the lattice, respectively.

Using the formula of Ernst and Binder[9], we have:

$$D_B = \phi_B - \frac{1}{4} \tag{2.4}$$

with

$$\phi_B = \frac{1}{2}\langle v_1|\frac{1}{1-\xi(1+\hat{T})}|v_1\rangle|_{\xi=1} \tag{2.5}$$

and $\langle v_1|$ the normalized 1-component of the velocity of the particle in the $\vec{e}_i$ basis:

$$\langle v_1| = \frac{1}{\sqrt{3}}(1, \frac{1}{2}, -\frac{1}{2}, -1, -\frac{1}{2}, \frac{1}{2}), \tag{2.6}$$

For the matrix $\hat{T}^{mirror}$, the six eigenvalues are:

$$\lambda_1 = 0, \lambda_2 = -2, \lambda_{3,5} = -1+\sqrt{\Gamma}, \lambda_{4,6} = -1-\sqrt{\Gamma} \tag{2.7}$$

with six corresponding orthonormal eigenvectors:

$$\begin{aligned}
\langle\lambda_1| &= \frac{1}{\sqrt{6}}(1,1,1,1,1,1) \\
\langle\lambda_2| &= \frac{1}{\sqrt{6}}(-1,1,-1,1,-1,1) \\
\langle\lambda_3| &= \frac{1}{\sqrt{3}}(\frac{2C_R-C_L}{2\sqrt{\Gamma}}, 1, \frac{2C_L-C_R}{2\sqrt{\Gamma}}, -\frac{1}{2}, -\frac{1}{2\sqrt{\Gamma}}, -\frac{1}{2}) \\
\langle\lambda_4| &= \frac{1}{\sqrt{3}}(\frac{C_L-2C_R}{2\sqrt{\Gamma}}, 1, \frac{C_R-2C_L}{2\sqrt{\Gamma}}, -\frac{1}{2}, -\frac{1}{2\sqrt{\Gamma}}, -\frac{1}{2}) \\
\langle\lambda_5| &= (\frac{C_L}{2\sqrt{\Gamma}}, 0, -\frac{C_R}{2\sqrt{\Gamma}}, -\frac{1}{2}, \frac{C_R-C_L}{2\sqrt{\Gamma}}, \frac{1}{2}) \\
\langle\lambda_6| &= (\frac{C_L}{2\sqrt{\Gamma}}, 0, -\frac{C_R}{2\sqrt{\Gamma}}, \frac{1}{2}, \frac{C_R-C_L}{2\sqrt{\Gamma}}, -\frac{1}{2})
\end{aligned} \tag{2.8}$$

where $\Gamma = C_L^2 - C_L C_R + C_R^2$. Using the projection operator $\sum_{i=1}^{6} |\lambda_i\rangle\langle\lambda_i|$ in (2.5), one obtains

$$\phi_B^{mirror} = \frac{1}{4C_L C_R} \tag{2.9}$$

while, similarly, for rotators one finds:

$$\phi_B^{rotator} = \frac{1}{2(C_L^2 + C_R^2)} \tag{2.10}$$

For the case of $C_L = C_R = \frac{1}{2}$, one has then

$$D_B = \frac{3}{4} \tag{2.11}$$

for both fixed and flipping mirrors as well as rotators.

## C. Computer Simulations

All the computer simulations were carried out on SiliconGraphics Indigo (SGI) with 32MB memory, SUN Sparc IPC with 24MB memory and VAX 3100 with 32MB memory. A virtual lattice of 90,000 × 90,000 sites was used. This lattice was conceptually divided into 300 × 300 blocks of 300 × 300 sites each. Two different arrays were used: one to record the position of a block, the other to record the position of the particle in the block. Actual memory was not assigned to a block (i.e., using "malloc" in C language) until the particle entered it. A flag is used to mark whether a block has been visited by the moving particle or not: if the block was not visited, i.e., the flag is 0, we assign memory to and put scatterers randomly on this block; otherwise, i.e., if the flag is 1, we continue to use the old configuration of scatterers that was on the block before. After the particle finished moving at the cut-off time step, we clear the memory and reset the flag for each block to 0.

The advantage of this scheme[5] is that only a small fraction of the memory for a 90,000 $\times$ 90,000 array is actually used, since we do not have to reserve memory for those areas that are never visited by a particle.

Another advantage of this memory allocation procedure appears when resetting the lattice to the blank condition, which must be done after each trajectory is completed. For a large array this operation itself would take a significant amount of time and would be required for each particle, even for very small closed orbits. In the present method, only the blocks that have been entered by the particle need to be reset when the trajectory is completed (i.e., using "free" in C language), significantly reducing the average time required for this operation.

About 30,000 independent particles, initially placed randomly on the lattice, were studied. In the case of flipping scatterers one particle was studied at a time. The calculations were done up to $2^{20}$ to $2^{26}$ time steps. The statistical errors were determined by doing the calculations in two steps: first an average was made over all 10,000 particles, with a different random configuration of the scatterers for each par-ticle, then further averages were computed over typically three runs, involving three samples of 10,000 particles for each. The standard deviations of the mean are plotted as the error bars of the data in the figures. If the error bar does not appear, the error bar is inside the symbol.

C.1 Fixed Scatterers

Like on the square and triangular lattices the diffusion is non-Gaussian since there are closed orbits on the honeycomb lattice for both the mirror and the rotator models.

a. $C_L = C_R = \frac{1}{2}$

In this case the two models behave identically, since they can be mapped into each other in a similar way as on the square lattice[10]. There is no-diffusion (Class IV), i.e., for increasing $t$ the diffusion coefficient $D(t)$ goes to zero, see fig.3a, means that the mean square displacement $\Delta(t)$ is bounded, i.e., there are no extended closed orbits; consequently, the distribution function $\hat{P}(r,t)$ does not correspond to that of a Gaussian diffusion process, but exhibits a sharp peak near the origin at $r_{max} \simeq 1-2$, and its shape does not appear to change any more after about $2^{21}$ time steps (cf.fig.3b). Since the behavior of the two models is the same, we show that for the rotator model only.

b. $C_L \neq C_R \, (C_L + C_R = 1)$

Like on the fully occupied square lattice, the rotator model on the honeycomb lattice behaves very differently from the mirror model. For, while the mirror model exhibits no-diffusion (Class IV) at least for the concentrations $0.35 \leq C_{L,R} \leq 0.65$, which we investigated (cf.fig.4a), the rotator model appears to possess two symmetric isolated critical points at $C_{L_{cr},R_{cr}} = 0.541$, $C_{R_{cr},L_{cr}} = 0.459$, where the diffusive behavior is anomalous (Class II). This differs from the results reported by Catalá et al.[11], who found a critical *line* from about $C_L = 0.541$ and $C_R = 0.459$ to $C_L = 0.459$ and $C_R = 0.541$, of which only the end points are consistent with our results. For all other concentrations, away from these critical concentrations, the diffusive behavior is no-diffusion (Class IV) (see figs.4b and 5). This behavior can also be seen in the difference in the number of open orbits as a function of time for $C_L = C_{L_{cr}}$ and $C_L \neq C_{L_{cr}}$ (and similarly for $C_{R_{cr}}$) (see fig.6a). For the rotator model in fig.6a, the number of open orbits $N_o(t)$ has a maximum value at $C_{L_{cr}}$ (or $C_{R_{cr}}$) and approaches a behavior as $t^{-1/7}$[1], i.e., we determined the size distribution exponent of open orbits characterized by $\tau = \frac{15}{7}$ from $N_o(t) \sim t^{2-\tau} = t^{-1/7}$. At the same time, we determined independently the fractal dimension $d_f = \frac{7}{4}$ from the product of probability of an open orbit $P_o(t)$ and the mean square displacement of a particle on an open orbit $\Delta_o(t)$ divided by $t$, i.e., from $P_o(t)\Delta_o(t)/t \sim t^{-1/7}t^{2/d_f}/t$ which appears to approach a constant (cf.fig.6c). This suggests that the hyperscaling relation $\tau - 1 = 2/d_f$ is valid at the critical points[1]. At all other concentrations, the orbits appear to close the quicker the further away $C_L$ ($C_R$) are from $C_{L_{cr}}$ ($C_{R_{cr}}$). Although we expect that at $C_{L_{cr}}$ (or $C_{R_{cr}}$), all orbits will close eventually, it may take an infinite number of time steps to do so, i.e., there exist extended closed orbits[1]. We should point out that it appears to take much longer for the diffusion process on the honeycomb lattice at a critical point to reach its asymptotic behavior than on the square and especially the triangular lattices. To what extent this is related to the small coordination number (3) of the honeycomb lattice is unclear. We are therefore not as certain about the existence of the two critical points and the hyperscaling relation for the honeycomb lattice, as we are for the other two lattices. In order to settle this point unambiguously we would have to extend our calculations far beyond our present maximum of $2^{26}$ time steps, where each data point in figs.4a, 6a and 6c typically already takes a few weeks on our SGI.

For the mirror model, no such critical points (cf.fig.6b) are found. In fact, the number of open orbits decreases gradually with increasing values of $C_L$ when $C_L$ varies from 0 to 0.5, where it is a minimum and similarly for $0 < C_R \leq 0.5$. This is because it is more difficult in the mirror model for a moving particle to make a closed orbit when there are more mirrors of one type than the other, due to an increased possibility of zig-zag motion (cf.fig.2c in ref.[1]).

C.2 Flipping Scatterers

a. $C_L = C_R = \dfrac{1}{2}$

Like on the fully occupied square and triangular lattices, the diffusive behavior of the mirror and the rotator models on the fully occupied honeycomb lattice is identical. This can be argued in the same way as for fixed scatterers at $C_L = C_R = \frac{1}{2}$. We find that the diffusion coefficient goes to zero and that the radial distribution function $\hat{P}(r,t)$ does not appear to change anymore after a number of time steps $t_{cr} \approx 2^{11}$, see fig.7a-b (in fig.7b, since the behavior of the two models is the same, we show that for the rotator model only), while the maximum of $\hat{P}(r,t)$ appears to remain fixed at a distance $r = r_{max} \approx 10$ lattice distances from the origin. We classify this diffusive behavior still as Class IV, since it is similar to that found for fixed scatterers in Section C.1, except that $r_{max}$ is much larger now. We observe that after about $2^{15}$ time steps virtually all particles are in closed orbits (see fig.8), suggesting that all trajectories will eventually become closed orbits. The smallest closed orbit can be considered to consist of two "reflectors" (i.e., the closed orbit consists of two parts connected by a single line, where each of these two parts is called a "reflector")[2] with a period of 92 time steps, where the particle suffers 92 collisions on 34 lattice sites (cf.fig.9). This closed orbit is much larger than the corresponding one of 6 lattice sites for fixed scatterers, which is the origin of the larger value of $r_{max}$ for the flipping

[2]L. Bunimovich and S. Troubetzkoy use a more general definition of a "reflector"[12].

scatterer model.

b. $C_L \neq C_R\,(C_L + C_R = 1)$

In this case, the diffusive behavior of both the rotator and the mirror model is Class IV, but it is not entirely identical for the two models.

For the flipping rotator model, the diffusion coefficient takes a shorter time for $C_L \neq C_R$ to vanish than for $C_L = C_R$ (see fig.7a). This is so, because for $C_L \neq C_R$, when one type of rotators is more numerous than the other, the closed orbits are less extended than when $C_L = C_R$ (cf.figs.10a-b), since for $C_L \neq C_R$ the tendency for the moving particle to bend in one direction and then return to its original position is larger than when an equal number of left and right rotators is present.

We remark that the closed orbits found in our simulations for $C_L \neq C_R$ appear to be of a different type than those observed for $C_L = C_R$, viz. without "reflectors". This difference is surely only apparently so, since for all $C_L$ and $C_R \neq 0$ closed orbits with or without "reflectors" will occur. However, their frequency of occurrence will depend very much on $C_L/C_R$. Thus, while for $C_L = C_R$ those with "reflectors" are the only ones seen in the simulations, for $C_L \neq C_R$ other types of orbits appear as well (see fig.10a). Similarly on the square lattice for flipping rotators, "reflectors" are not necessary for closed orbits either (cf.figs.10c-d). However, on the triangular lattice, for both flipping rotators and mirrors, closed orbits never contain "reflectors" (as defined above) (see figs.10e-f). This can be understood as follows: for a “reflector” to exist, it is necessary that the the moving particle goes through at least one bond at least once in opposite directions, i.e., the moving particle must turn over an angle of an odd number times of $\pi$, consistent with the scattering rules. However, the latter is not possible for the triangular lattice because an angle of an odd number times of $\pi$ can not be produced by any combination of the possible scattering angles $\pm 2\pi/3$ and 0.

For the flipping mirror model, the diffusion coefficient goes to zero for both $C_L \neq C_R$ and $C_L = C_R$. However, for this model it takes a longer time for the diffusion coefficient to vanish when $C_L \neq C_R$ than for $C_L = C_R$, see fig.7a, since there exists now a tendency for the particle to propagate (zig-zag motion) when the particle hits a region with more right (left) mirrors than left (right) mirrors.

# 3 Quasi-lattice

A. Introduction

The quasi-lattice we studied is a Fibonacci lattice rather than a Penrose lattice, since the former gives a higher density of vertices (for a given lattice size). It is shown in fig.11. It was constructed in the following way[13]. First a “star” of 5 two-dimensional vectors $\vec{s}_1$, $\vec{s}_2$, $\vec{s}_3$, $\vec{s}_4$, $\vec{s}_5$ is drawn (cf.fig.12a). This “star” has pentagonal orientational symmetry, i.e., the angle between each pair of adjacent vectors is $2\pi/5$. Next a grid, i.e., a set of, in principle, infinite, but here in fact 73, quasi-periodically spaced parallel lines are introduced perpendicular to each star vector $\vec{s}_i (i = 1, ..., 5)$ with spacings:

$$x_{n,i} = \vec{r}_{n,i} \cdot \vec{s}_i = T_i[n + \alpha_i + \frac{1}{\rho_i}\lfloor\frac{n}{\sigma_i} + \beta_i\rfloor] \quad (3.1)$$

Here $x_{n,i}$ is the distance along the direction $\vec{s}_i$ $(i = 1, ..., 5)$ between the origin and the n-th parallel line; $\vec{r}_{n,i}$ is a point on the n-th parallel line perpendicular to $\vec{s}_i (i = 1, ..., 5)$; $\lfloor f \rfloor$ is the floor function, denoting the integer part of $f$ and $n = 1, 2, ...., N$ (here $N$ is 73). These 5 grids compose the grid-space. Furthermore, $T_i, \alpha_i, \rho_i, \sigma_i$ and $\beta_i$ are constant parameters, of which the $\sigma_i$ must be irrational real numbers in order to get non-periodic spacings[3]. Eq.(3.1) defines a quasi-periodic "sequence of intervals" such that the intervals $\Delta x_i = x_{n,i} - x_{n-1,i}$ between two adjacent parallel lines have the property:

$$\Delta x_i = \begin{cases} T_i & \text{if } \lfloor n/\sigma_i + \beta \rfloor - \lfloor (n-1)/\sigma_i + \beta \rfloor = 0, \\ T_i(1 + 1/\rho_i) & \text{if } \lfloor n/\sigma_i + \beta \rfloor - \lfloor (n-1)/\sigma_i + \beta \rfloor = 1. \end{cases} \quad (3.2)$$

That is, there are only two possible intervals between adjacent parallel lines for each grid, $T_i(1 + 1/\rho_i)$ and $T_i$, which appear in a quasi-periodic sequence, where the ratio of the number of $T_i(1 + 1/\rho_i)$ distances to the number of $T_i$ distances equals $1/(1 - \sigma_i)$. Thus, the parameter $\sigma_i$ determines the relative frequencies of the two different spacings in the sequence and $\rho_i$ determines the ratio of the two spacing lengths that occur between the lines in each grid. In our calculations we took the parameters $\rho_i = \sigma_i \equiv \tau = (\sqrt{5} + 1)/2$ $(i = 1, ..., 5)$, the golden mean and we chose $T_i = 0.25, 0.8, 0.25, 0.7, 0.7$; $\alpha_i = -30.0, -30.0, -31.0, -72.0, -71.0$ and $\beta_i = 0.7, 0.8, 1.1, 1.2, 0.9$, for $i = 1, ..., 5$, respectively, in order to make 5 grids that overlapped as much as possible. In this way a maximum number density of vertices of the quasi-lattice was obtained. The grid thus generated is called a Fibonacci pentagrid[14], to which we restrict ourselves here. The effect of a change in $\alpha_i$ is simply to translate each entire grid in the direction $\vec{s}_i$ over a distance $T_i\alpha_i$, whereas a change in $\beta_i$ alters the sequence of long and short spacings. In each grid, each line normal to the star vector $\vec{s}_i$ is labeled by an integer $k_i$ which represents the location of its position along the $\vec{s}_i$ direction. The lines divide the grid space into non-intersecting open regions through which no lines pass (the regions can be arbitrarily small). Each such region is specified (uniquely) by $M$ (here 5) integers $(k_1, k_2, ..., k_M)$: if $\vec{x}_0$ is any point in the region, then $k_i$ is the label of the line normal to $\vec{s}_i$ such that $\vec{x}_0$ lies between the lines labeled by $k_i$ and $k_i + 1$. Finally the generalized dual method[13] is applied to find the dual of the pentagrid. The "dual" is constructed by mapping each open region in grid space into a point $\vec{t} = \sum_{i=1}^{M} k_i \vec{s}_i$, which lies in a two dimensional space that we shall call the "cell-space". The points $\vec{t}$ are the vertices of a packing of the quasi-lattice by unit cells (rhombuses). So the dual transformation maps a grid-space into a cell-space such that open regions in grid-space are mapped into points in cell-space and the points in grid-space are mapped into the open regions (rhombuses) in cell-space. This gives a tiling of the plane with two differently shaped rhombuses (one is a thin rhombus with angles $36^o, 144^o, 36^o$ and $144^o$, the other is a fat rhombus with angles $72^o, 108^o, 72^o$ and $108^o$) and produces the quasi-lattice shown in fig.11. The lengths of all bonds of the quasi-lattice are equal and are chosen to be unity. Since our quasi-lattice has no translational or rotational symmetries, we define two lattice sites as the same, if one lattice site can be mapped into the other by a pure translation. The Fibonacci quasi-lattice has 658 different lattice sites[4]. If each lattice site were to appear on the lattice with the same probability, i.e., 1/658, then the average coordination number would be 5.6839. Since in our lattice, not every lattice site appears with the same probability the average coordination number is actually about 3.9887.

[3]For, if we choose for $\sigma_i$ in eq.(3.1) rational real numbers, e.g., $\sigma_i = p_i/q_i$ with both $p_i$ and $q_i$ positive integers, we will get periodic spacings, $x_{m,i} = T'_i(m + \gamma_i)$ and therefore a periodic lattice with period $T'_i$. Here $m$ is an integer $m = 1, 2, ..., \lfloor N/p_i \rfloor$, where each $m$ stands for a set of $p_i$ adjacent lines in the sequence of $N$ lines, $T'_i = T_i(p_i + q_i/\rho_i)$ and $\gamma_i = (\alpha_i + \lfloor \beta_i \rfloor / \rho_i)/(p_i + q_i/\rho_i)$.

B. Equations of Motion

[4]The lattice sites can differ in three ways: a) different number of bonds to the nearest neighbors; b) same number of bonds to the nearest neighbors, but different angles between adjacent bonds; c) same number of bonds to the nearest neighbors and same angles between adjacent bonds, but different orientation of these bonds.

We restrict ourselves to the rotator model, where the particle will turn to its left (right) over the largest available angle between lattice bonds if there is a left (right) rotator (cf.fig.12c). This leads to the following equations of motion for both fixed and flipping rotators:

$$n_i(\vec{r}+\vec{e}_i,t+1) = [n_{i+1}(\vec{r},t)I^{i+1\,i}(\vec{r}) + n_{i+2}(\vec{r},t)I^{i+2\,i}(\vec{r}) +$$
$$n_{i+3}(\vec{r},t)I^{i+3\,i}(\vec{r}) + n_{i+4}(\vec{r},t)I^{i+4\,i}(\vec{r})]m_R(\vec{r},t) + [n_{i+6}(\vec{r},t)I^{i+6\,i}(\vec{r}) +$$
$$n_{i+7}(\vec{r},t)I^{i+7\,i}(\vec{r}) + n_{i+8}(\vec{r},t)I^{i+8\,i}(\vec{r}) + n_{i+9}(\vec{r},t)I^{i+9\,i}(\vec{r})]m_L(\vec{r},t) \quad (3.3)$$

$$(i = 1, ..., 10; \text{ mod } 10)$$

Here $\vec{e}_i$ is the unit vector defining the velocity direction $i$ $(i = 1, ..., 10)$ and $n_i(\vec{r},t)$, $m_R(\vec{r},t)$ and $m_L(\vec{r},t)$ have the same meaning as in the equations for the honeycomb lattice. $I^{ij}(\vec{r})$ is a geometric factor, which can only have the values 1 or 0, depending on whether the particle can or cannot change from $\vec{e}_i$ to $\vec{e}_j$, respectively, at $\vec{r}$. In the square brackets before $m_R(\vec{r},t)$ or $m_L(\vec{r},t)$ in eq.(3.3), only one $I^{ij}(\vec{r})$ is $\neq 0$, viz. that one which allows the particle coming in with velocity direction $i$ to turn to the allowed direction $j$ to its right or left, respectively.

## C. Boltzmann Approximation

A Boltzmann approximation to the eq.(3.3) is obtained by averaging both sides of the equation over all possible random configurations of the scatterers and the particle and ignoring any correlations between scatterer occupation and particle velocity at each lattice site. One obtains then the equations:

$$f_i(t+1) = f_i(t) + \sum_{j=1}^{10} T_{ij} f_j(t) \quad (i = 1, ..., 10) \tag{3.4}$$

where $f_i(t)$ is the probability to find a particle with the velocity direction along $\vec{e}_i$ at time $t$. $T_{ij}$ are the elements of a 10 x 10 collision matrix $\hat{T}$, with off-diagonal matrix elements proportional to $C_L$ or $C_R$. For $C_L = C_R = \frac{1}{2}$, to which we restrict ourselves here, $\hat{T}$ reads:

$$\hat{T} = \frac{1}{2}\begin{pmatrix}
-2 & <I^{2\,1}> & <I^{3\,1}> & <I^{4\,1}> & <I^{5\,1}> & 0 & <I^{7\,1}> & <I^{8\,1}> & <I^{9\,1}> & <I^{10\,1}> \\
<I^{1\,2}> & -2 & <I^{3\,2}> & <I^{4\,2}> & <I^{5\,2}> & <I^{6\,2}> & 0 & <I^{8\,2}> & <I^{9\,2}> & <I^{10\,2}> \\
<I^{1\,3}> & <I^{2\,3}> & -2 & <I^{4\,3}> & <I^{5\,3}> & <I^{6\,3}> & <I^{7\,3}> & 0 & <I^{9\,3}> & <I^{10\,3}> \\
<I^{1\,4}> & <I^{2\,4}> & <I^{3\,4}> & -2 & <I^{5\,4}> & <I^{6\,4}> & <I^{7\,4}> & <I^{8\,4}> & 0 & <I^{10\,4}> \\
<I^{1\,5}> & <I^{2\,5}> & <I^{3\,5}> & <I^{4\,5}> & -2 & <I^{6\,5}> & <I^{7\,5}> & <I^{8\,5}> & <I^{9\,5}> & 0 \\
0 & <I^{2\,6}> & <I^{3\,6}> & <I^{4\,6}> & <I^{5\,6}> & -2 & <I^{7\,6}> & <I^{8\,6}> & <I^{9\,6}> & <I^{10\,6}> \\
<I^{1\,7}> & 0 & <I^{3\,7}> & <I^{4\,7}> & <I^{5\,7}> & <I^{6\,7}> & -2 & <I^{8\,7}> & <I^{9\,7}> & <I^{10\,7}> \\
<I^{1\,8}> & <I^{2\,8}> & 0 & <I^{4\,8}> & <I^{5\,8}> & <I^{6\,8}> & <I^{7\,8}> & -2 & <I^{9\,8}> & <I^{10\,8}> \\
<I^{1\,9}> & <I^{2\,9}> & <I^{3\,9}> & 0 & <I^{5\,9}> & <I^{6\,9}> & <I^{7\,9}> & <I^{8\,9}> & -2 & <I^{10\,9}> \\
<I^{1\,10}> & <I^{2\,10}> & <I^{3\,10}> & <I^{4\,10}> & 0 & <I^{6\,10}> & <I^{7\,10}> & <I^{8\,10}> & <I^{9\,10}> & -2
\end{pmatrix} \tag{3.5}$$

Here $< I^{ij} >$ is the fraction of all lattice sites for which $I^{ij}(\vec{r})$ is not zero. The values of $< I^{ij} >$ for the Fibonacci lattice are given in the Appendix. Using the method of Ernst and Binder[9] as above (cf.eqs.(2.4 - 2.5)) with the $T_{ij}$ of eq.(3.5), the diffusion coefficient in the Boltzmann approximation can be obtained. Since the diffusion tensor is not isotropic, we use for the Boltzmann diffusion coefficient $D_B = \frac{1}{2}(D_{xx,B} + D_{yy,B})$. Using that $\langle v_x| = \frac{1}{\sqrt{5}}(1, \cos\frac{\pi}{5}, \cos\frac{2\pi}{5}, -\cos\frac{2\pi}{5}, -\cos\frac{\pi}{5}, -1, -\cos\frac{\pi}{5}, -\cos\frac{2\pi}{5}, \cos\frac{2\pi}{5}, \cos\frac{\pi}{5})$ and $\langle v_y| = \frac{1}{\sqrt{5}}(0, \cos\frac{2\pi}{5}, \cos\frac{\pi}{5}, \cos\frac{\pi}{5}, \cos\frac{2\pi}{5}, 0, -\cos\frac{2\pi}{5}, -\cos\frac{\pi}{5}, -\cos\frac{\pi}{5}, -\cos\frac{2\pi}{5})$, which are the $x$ and $y$ components of the velocity of the particle in the $\vec{e}_i$ $(i = 1, ..., 10)$ basis, respectively (cf.eq.(2.6)), leads then with eqs.(2.4) and (2.5) to a value $D_B \simeq 0.165$. This value can be compared with the value $D_B \simeq 0.1661 \pm 0.0005$ as determined by computer simulation for a probabilistic model with a scattering rule of equal probability for the particle to scatter over the largest angle to its left or to its right.

## D. Computer Simulations

A unit cell of 22,350 lattice sites and Ziff et al.'s[5] method described above for the honeycomb lattice were used. Since it is not possible here to impose strictly periodic boundary conditions, we chose the following quasi-periodic boundary conditions: each time that a particle leaves the unit cell along a bond with a given velocity at one side, it reappears on the opposite side with the same velocity direction along the bond closest to where it would have reappeared for periodic boundary conditions. About 30,000 independent particles were used and the calculations were pursued up to $2^{20}$ time steps. For flipping scatterers one particle was studied at a time. As explained for the honeycomb lattice, the standard deviations of the mean are plotted as the error bars of the data in the figures.

### D.1 Fixed Scatterers

In this case, we found no-diffusion (Class IV) everywhere, since all orbits clearly seemed to be closed after a finite number of time steps. Therefore, unlike in the honeycomb lattice, there appear to be no critical points. As a consequence, the mean square displacement is bounded, as can be seen this both from the diffusion coefficient and the number of closed orbits (see figs.13a-b). We also checked the effect of boundary conditions on the computer simulation results and found virtually no difference in the diffusive behavior between a basic block with 22,350 or 10,940 sites (cf.fig.13c).

D.2 Flipping Scatterers

Like on the honeycomb lattice, the diffusion coefficient goes to zero, while the distribution function $\hat{P}(r,t)$ exhibits a sharp maximum at $r = r_{max} \approx 10$ lattice sites and does not appear to change anymore after $2^{11}$ time steps (cf.figs.14a-b). It seems that all orbits will eventually be closed. Because of the irregularity of the lattice we have not been able to identify yet the smallest closed orbit; figs.15a-b give two examples of closed orbits.

# 4 Discussion

1. We summarize the diffusive behavior for the fixed and flipping scatterer models for $C_L$ or $C_R \neq 0$ on all four lattices investigated here as well as in ref.[1][5], in Table

[5]For the diffusive behavior for $C = C_L$ (or $C = C_R$), i.e., a lattice covered by only one type of scatterers, see ref.[6].

I and II, respectively.

The two tables reflect how the diffusion process differs on the fully occupied honeycomb and quasi-lattices from those studied before and enable us to get a sense of the influence of the lattice structure as well as of the scattering rules on the diffusion process of a particle on these lattices. We have the following comments.

a) In Table I critical lines or points for the fixed rotator model occur for all lattices except the quasi-lattice. For the mirror model on the square lattice, the critical behavior occurs for all concentrations along the critical line $C = 1$. In Table II, a phase transition occurs at $C = 1$ for the rotator model on the square lattice as well as for both the rotator and mirror models on the triangular lattice.

One can interpret the results in Tables I and II in terms of dynamical analogues of phase transitions, critical points and critical lines. A transition from Class IV to Class II exhibits some analogy to a second order phase transition, in that the behavior implies the appearance of extended closed orbits (cf. long range correlations at critical points or on critical lines) when before trapping, i.e., only short closed orbits, occurred (cf. short range correlations away from critical points or lines). On the other hand, the less subtle transitions from Class II to super-diffusion or Class IV to normal diffusion (Class I) or quasi-normal diffusion to propagation bear more resemblance to a first order phase transition. It is not clear whether this analogy can be pushed beyond this descriptive stage.

b) The triangular lattice is an exception in that the rotator and mirror models always behave the same for any concentration of scatterers for both fixed[1] and flipping scatterers for the same reason as argued in ref.[1] for fixed scatterers.

c) In Tables I and II, the diffusive behavior of the moving particle covers a wide variety of different cases: propagation, super-diffusion, normal, quasi-normal, anomalous and no-diffusion.

2. We note that the quasi-normal behavior for $0 < C < 1$ in the triangular lattice listed in Table II, differs from the normal behavior mentioned in ref.[2] before, because the finite probability for closed orbits to occur was overlooked in this paper[6]. This makes the diffusion process strictly speaking non-Gaussian[10]. However, the fraction of closed orbits is so small (see fig.16a) that for all practical purposes the diffusion can be considered Gaussian; as a consequence we call it quasi-normal (cf.fig.16b). We conjecture that a similar situation may obtain in $d = 3$ for the simple cubic lattice, for instance, where closed orbits will occur relatively frequently for short times when the particle hovers near the origin, but where, with increasing time, return to the origin, once the particle is away, will become increasingly difficult.

3. We note that the mirror model for fixed scatterers is time reversal invariant

[6]This was noticed by L. Bunimovich and S. Troubetzkoy (see ref.[10]) and independently by one of us (F. W., unpublished).

on the square and the triangular lattices but not on the honeycomb lattice. This is due to the small scattering angle ($\pi/3$) of the particle on the honeycomb lattice as compared to those on the square and the triangular lattices, where they are $\pi/2$ and $2\pi/3$, respectively. However, the flipping mirror model on the honeycomb lattice is time reversal invariant, while this is not so on the other two lattices.

4. As to the nature of closed orbits for flipping scatterers, empty sites are necessary on the square and triangular lattices, but not on the honeycomb and quasi-lattices ($C = 1$). "Reflectors" occur but are not necessary for closed orbits on the square, honeycomb and quasi-lattices, i.e., closed orbits may or may not include "reflectors", while on the triangular lattice, there are no "reflectors" (cf.figs.10a-f).

5. Even when all orbits close, there are for all lattices a number of different cases to distinguish of how this closing occurs.

($i$). Many trajectories close very near the origin (where $\hat{P}$ has a maximum at $r = r_{max} \approx 1 - 2$) but most trajectories are extended and close gradually when $t \to \infty$, as shown in the slow approach of $\hat{P}(r,t)$ to its asymptotic shape (cf.figs.2a and 7a in ref.[1]). This obtains in the case of anomalous diffusion (Class II) for fixed scatterers. Then the decrease in the number of open orbits (characterized by $\tau$) and the mean square displacement (characterized by $d_f$) are connected by the hyperscaling relation $\tau - 1 = 2/d_f$.

($ii$). Almost all orbits close very near the origin at $r = r_{max} \approx 1-2$ and $\hat{P}(r,t)$ is virtually stationary for $t > t_{cr}$ (cf.fig.3b and fig.11b in ref.[1]). This obtains in the case of no-diffusion (Class IV) for fixed scatterers. All orbits close after a finite number of time steps.

($iii$). Almost all orbits close near the origin at $r \approx r_{max} \approx 10$ and $\hat{P}(r,t)$ is virtually stationary for $t > t_{cr} \approx 2^{11}$ (cf.figs.7b and 14b). This obtains in the case of no-diffusion (Class IV) for flipping scatterers. All orbits close after a finite number of time steps.

($iv$). Only a small fraction of trajectories close near the origin, yet $\hat{P}(r,t)$ has a maximum at $r_{max} \simeq 1-2$ for all sufficiently large $t$. The particle appears to move overwhelmingly in unbounded trajectories (cf.fig.16b). This obtains in the case of quasi-normal diffusion for both flipping rotators and mirrors on the triangular lattice for $C < 1$.

($v$). All trajectories will close eventually with only a few closed orbits near the origin, while the overwhelming majority carry particles whose mean square displacement grows faster than $t$ (cf. fig.7b in ref.[1]). This obtains in the case of super-diffusion for $0 < C < 1$ for the fixed mirror model on the square lattice. The probability to find a closed orbit decreases as $\sim 1/\ln t$[1].

6. Although the theorems of Bunimovich and Troubetzkoy rigorously establish

some of the results found in these computer simulations on the square and the triangular lattices and we were able to present physical arguments for a number of the observed behaviors, clearly a deeper understanding of the diffusion process of a particle on a lattice in its dependence on lattice structure and scattering rules is lacking.

## Acknowledgements

The authors are grateful to Prof. L. A. Bunimovich and Drs. C. H. Chou, D. J. Callaway, A. Hof and M. S. Cao for helpful remarks. Part of this work was supported by Department of Energy under grant No. DE-FG02-88ER13847.

## Appendix:

$$< I^{1\,1} >=< I^{6\,6} >= 0 \qquad < I^{1\,2} >=< I^{7\,6} >= 6.7338436198042825E-03$$
$$< I^{1\,3} >=< I^{8\,6} >= 2.9648289894390078E-02 \qquad < I^{1\,4} >=< I^{9\,6} >= 3.1925201046410232E-02$$
$$< I^{1\,5} >=< I^{10\,6} >= 8.1387462455188451E-03 \qquad < I^{1\,6} >= 0$$
$$< I^{1\,7} >=< I^{2\,6} >= 7.4120724735975196E-03 \qquad < I^{1\,8} >=< I^{3\,6} >= 3.0907857765720376E-02$$
$$< I^{1\,9} >=< I^{4\,6} >= 2.8679391531828311E-02 \qquad < I^{1\,10} >=< I^{5\,6} >= 6.2978393566514873E-03$$
$$< I^{2\,1} >=< I^{6\,7} >= 6.4431741110357524E-03 \qquad < I^{2\,2} >=< I^{7\,7} >= 0$$
$$< I^{2\,3} >=< I^{8\,7} >= 2.0395310531925201E-02 \qquad < I^{2\,4} >=< I^{9\,7} >= 3.4202112198430385E-02$$
$$< I^{2\,5} >=< I^{10\,7} >= 1.3613021993992830E-02 \qquad < I^{2\,6} >=< I^{1\,7} >= 7.4120724735975196E-03$$
$$< I^{2\,7} >= 0 \qquad < I^{2\,8} >=< I^{3\,7} >= 2.5336692180990214E-02$$
$$< I^{2\,9} >=< I^{4\,7} >= 3.5267900397248328E-02 \qquad < I^{2\,10} >=< I^{5\,7} >= 1.2159674450150179E-02$$

$< I^{3\,1} >=< I^{6\,8} >= 2.8727836449956399E-02$ $< I^{3\,2} >=< I^{7\,8} >= 2.1461098730743145E-02$
$< I^{3\,3} >=< I^{8\,8} >= 0$ $< I^{3\,4} >=< I^{9\,8} >= 6.2009495203953106E-02$
$< I^{3\,5} >=< I^{10\,8} >= 3.6043019087297743E-02$ $< I^{3\,6} >=< I^{1\,8} >= 3.0907857765720376E-02$
$< I^{3\,7} >=< I^{2\,8} >= 2.5336692180990214E-02$ $< I^{3\,8} >= 0$
$< I^{3\,9} >=< I^{4\,8} >= 6.2397054548977812E-02$ $< I^{3\,10} >=< I^{5\,8} >= 3.4831896134095534E-02$
$< I^{4\,1} >=< I^{6\,9} >= 2.9938959403158609E-02$ $< I^{4\,2} >=< I^{7\,9} >= 3.4686561379711268E-02$
$< I^{4\,3} >=< I^{8\,9} >= 5.7601007654297065E-02$ $< I^{4\,4} >=< I^{9\,9} >= 0$
$< I^{4\,5} >=< I^{10\,9} >= 2.1121984303846526E-02$ $< I^{4\,6} >=< I^{1\,9} >= 2.8679391531828311E-02$
$< I^{4\,7} >=< I^{2\,9} >= 3.5267900397248328E-02$ $< I^{4\,8} >=< I^{3\,9} >= 6.2397054548977812E-02$
$< I^{4\,9} >= 0$ $< I^{4\,10} >=< I^{5\,9} >= 2.4658463327196977E-02$
$< I^{5\,1} >=< I^{6\,10} >= 8.1871911636469334E-03$ $< I^{5\,2} >=< I^{7\,10} >= 1.3661466912120918E-02$
$< I^{5\,3} >=< I^{8\,10} >= 3.5703904660401124E-02$ $< I^{5\,4} >=< I^{9\,10} >= 2.1073539385718438E-02$
$< I^{5\,5} >=< I^{10\,10} >= 0$ $< I^{5\,6} >=< I^{1\,10} >= 6.2978393566514873E-03$
$< I^{5\,7} >=< I^{2\,10} >= 1.2159674450150179E-02$ $< I^{5\,8} >=< I^{3\,10} >= 3.4831896134095534E-02$
$< I^{5\,9} >=< I^{4\,10} >= 2.4658463327196977E-02$ $< I^{5\,10} >= 0$
$< I^{6\,1} >= 0$ $< I^{6\,2} >=< I^{7\,1} >= 7.7511869004941382E-03$
$< I^{6\,3} >=< I^{8\,1} >= 3.1150082356360818E-02$ $< I^{6\,4} >=< I^{9\,1} >= 2.9115395794981107E-02$
$< I^{6\,5} >=< I^{10\,1} >= 6.6853987016761942E-03$ $< I^{6\,6} >=< I^{1\,1} >= 0$
$< I^{6\,7} >=< I^{2\,1} >= 6.4431741110357524E-03$ $< I^{6\,8} >=< I^{3\,1} >= 2.8727836449956399E-02$
$< I^{6\,9} >=< I^{4\,1} >= 2.9938959403158609E-02$ $< I^{6\,10} >=< I^{5\,1} >= 8.1871911636469334E-03$
$< I^{7\,1} >=< I^{6\,2} >= 7.7511869004941382E-03$ $< I^{7\,2} >= 0$
$< I^{7\,3} >=< I^{8\,2} >= 2.3883344637147563E-02$ $< I^{7\,4} >=< I^{9\,2} >= 3.4395891870942738E-02$
$< I^{7\,5} >=< I^{10\,2} >= 1.1723670186997384E-02$ $< I^{7\,6} >=< I^{1\,2} >= 6.7338436198042825E-03$
$< I^{7\,7} >=< I^{2\,2} >= 0$ $< I^{7\,8} >=< I^{3\,2} >= 2.1461098730743145E-02$
$< I^{7\,9} >=< I^{4\,2} >= 3.4686561379711268E-02$ $< I^{7\,10} >=< I^{5\,2} >= 1.3661466912120918E-02$
$< I^{8\,1} >=< I^{6\,3} >= 3.1150082356360818E-02$ $< I^{8\,2} >=< I^{7\,3} >= 2.3883344637147563E-02$
$< I^{8\,3} >= 0$ $< I^{8\,4} >=< I^{9\,3} >= 6.3317507993411491E-02$
$< I^{8\,5} >=< I^{10\,3} >= 3.4735006297839357E-02$ $< I^{8\,6} >=< I^{1\,3} >= 2.9648289894390078E-02$
$< I^{8\,7} >=< I^{2\,3} >= 2.0395310531925201E-02$ $< I^{8\,8} >=< I^{3\,3} >= 0$
$< I^{8\,9} >=< I^{4\,3} >= 5.7601007654297065E-02$ $< I^{8\,10} >=< I^{5\,3} >= 3.5703904660401124E-02$
$< I^{9\,1} >=< I^{6\,4} >= 2.9115395794981107E-02$ $< I^{9\,2} >=< I^{7\,4} >= 3.4395891870942738E-02$
$< I^{9\,3} >=< I^{8\,4} >= 6.3317507993411491E-02$ $< I^{9\,4} >= 0$
$< I^{9\,5} >=< I^{10\,4} >= 2.2526886929561089E-02$ $< I^{9\,6} >=< I^{1\,4} >= 3.1925201046410232E-02$
$< I^{9\,7} >=< I^{2\,4} >= 3.4202112198430385E-02$ $< I^{9\,8} >=< I^{3\,4} >= 6.2009495203953106E-02$
$< I^{9\,9} >=< I^{4\,4} >= 0$ $< I^{9\,10} >=< I^{5\,4} >= 2.1073539385718438E-02$

$< I^{10\,1} >=< I^{6\,5} >= 6.6853987016761942E-03$ $< I^{10\,2} >=< I^{7\,5} >= 1.1723670186997384E-02$
$< I^{10\,3} >=< I^{8\,5} >= 3.4735006297839357E-02$ $< I^{10\,4} >=< I^{9\,5} >= 2.2526886929561089E-02$
$< I^{10\,5} >= 0$ $< I^{10\,6} >=< I^{1\,5} >= 8.1387462455188451E-03$
$< I^{10\,7} >=< I^{2\,5} >= 1.3613021993992830E-02$ $< I^{10\,8} >=< I^{3\,5} >= 3.6043019087297743E-02$
$< I^{10\,9} >=< I^{4\,5} >= 2.1121984303846526E-02$ $< I^{10\,10} >=< I^{5\,5} >= 0$

# Figure Captions

Fig.1. Part of a honeycomb lattice.

Fig. 2. (a) The 6 velocity directions on the honeycomb lattice; after rotation over $\pi$, lattice site I transforms into lattice site II and vice versa; (b) examples of rotator and (c) of mirror scatterers, respectively.

Fig. 3. (a) Diffusion coefficient $D$ as a function of the time $t$ on a $\log_{10}$-$\log_{10}$ scale for the fixed rotator model on the honeycomb lattice for $C_L = C_R = 0.5$; (b) corresponding radial distribution function $\hat{P}(r,t)$ as a function of distance $r$ from the origin at $t = 2^{11}$ ($\diamond$), $t = 2^{21}$ (+) and $t = 2^{24}$ ($\square$), respectively.

Fig. 4. (a) Diffusion coefficient $D$ as a function of the time $t$ on a $\log_{10}$-$\log_{10}$ scale for the fixed mirror model on the honeycomb lattice for $C_L = 0.55$, $C_R = 0.45$ ($\diamond$), $C_L = 0.6$, $C_R = 0.4$ (+) and $C_L = 0.65$, $C_R = 0.35$ ($\square$); (b) Diffusion coefficient $D$ as a function of the time $t$ on a $\log_{10}$-$\log_{10}$ scale for the fixed rotator model on the honeycomb lattice for $C_L = C_R = 0.5$ ($\diamond$), $\mathbf{C_L = 0.541}$, $\mathbf{C_R = 0.459}$ ($\square$) $C_L = 0.55$, $C_R = 0.45$ (+) and $C_L = 0.6$, $C_R = 0.4$ ($\times$); critical concentration is printed in bold face.

Fig. 5. Phase diagram for the fixed rotator model on the fully occupied honeycomb lattice.

Fig. 6. Number of open orbits out of 10,000 trajectories on the honeycomb lattice as a function of $t$ on a $\log_2$-$\log_2$ scale for (a) fixed rotator model: $C_L = C_R = 0.5$ ($\diamond$), $C_L = 0.53$, $C_R = 0.47$ (+), $\mathbf{C_L = 0.541}$, $\mathbf{C_R = 0.459}$ ($\square$), $C_L = 0.55$, $C_R = 0.45$ ($\times$), $C_L = 0.57$, $C_R = 0.43$ ($\triangle$) and $C_L = 0.6$, $C_R = 0.4$ ($*$); (b) fixed mirror model: $C_L = C_R = 0.5$ ($\diamond$), $C_L = 0.55$, $C_R = 0.45$ (+), $C_L = 0.6$, $C_R = 0.4$ ($\square$), $C_L = 0.65$, $C_R = 0.35$ ($\times$); (c) contribution of diffusion coefficient from open orbits $P_o(t)\Delta_o(t)/t$ as a function of time $t$ on a $\log_2$-$\log_2$ scale for fixed rotator model on the honeycomb lattice for $\mathbf{C_L = 0.541}$, $\mathbf{C_R = 0.459}$ ($\diamond$), $C_L = 0.53$, $C_R = 0.47$ (+) and $C_L = 0.55$, $C_R = 0.45$ ($\square$).

Fig. 7. (a) Diffusion coefficient $D$ as a function of the time $t$ on a $\log_{10}$-$\log_{10}$ scale for the flipping rotator and mirror models on the honeycomb lattice for $C_L = C_R = 0.5$ for flipping rotator ($\diamond$) and flipping mirror (+), $C_L = 0.8$ and $C_R = 0.2$ for flipping rotator ($\square$) and flipping mirror ($\times$); (b) corresponding $\hat{P}(r,t)$ as a function of $r$ for the flipping rotator model on the honeycomb lattice for $C_L = C_R = 0.5$ at $t = 2^{15}$ ($\diamond$), $t = 2^{17}$ (+) and $t = 2^{19}$ ($\square$), respectively.

Fig. 8. Fraction of closed orbits on the honeycomb lattice for the flipping rotator model ($\diamond$) and flipping mirror model (+) (indistinguishable) for $C_L = C_R = 0.5$, as a function of $\log_2 t$.

Fig. 9. An example of the smallest closed orbit for the flipping rotator model on the honeycomb lattice with period of 92 time steps on 34 lattice sites with two “reflectors”: A and B. Up-triangles stand for right rotators and down-triangles for left rotators, respectively.

Fig. 10. A few typical examples of closed orbits for the flipping rotator model on different lattices. Up-triangles stand for right rotators; down-triangles for left rotators and squares for empty sites, respectively. (a) Period of 2209 time steps on 256 lattice sites without “reflectors” and empty sites for $C_L = 0.95$, $C_R = 0.05$ on the honeycomb lattice; (b) period of 852 time steps on 182 lattice sites with two “reflectors” (A and B) but without empty sites for $C_L = C_R = 0.5$ on the honeycomb lattice; it has a shorter period, but is more extended than that in fig.10(a) for $C_L \neq C_R$; (c) period of 2697 time steps on 232 lattice sites without “reflectors” but with empty sites for $C_L = 0.95$, $C_R = 0.01$ on the square lattice; (d) period of 337 time steps on 61 lattice sites with two “reflectors” (A and B) and empty sites for $C_L = 0.95$, $C_R = 0.01$ on the square lattice; (e) the smallest closed orbit of period of 18 time steps on 7 lattice sites without “reflectors” but with one empty site on the triangular lattice; (f) period of 36 time steps on 13 lattice sites without “reflectors” but with empty sites for $C < 1.0$ on the triangular lattice.

Fig. 11. Part of a Fibonacci quasi-lattice.

Fig. 12. (a) Five-“star” with pentagonal orientation symmetry; (b) the 10 velocity

directions on the Fibonacci quasi-lattice; (c) rotator scatterers.

Fig. 13. (a) Diffusion coefficient $D$ as a function of the time $t$ on a $\log_{10}$-$\log_{10}$ scale for the fixed rotator model on a Fibonacci quasi-lattice for $C_L = C_R = 0.5$ ($\diamond$), $C_L = 0.57$, $C_R = 0.43$ ($+$) and $C_L = 0.55$, $C_R = 0.45$ ($\Box$); (b) as in (a) the number of closed orbits for $C_L = C_R = 0.5$ ($\diamond$), $C_L = 0.51$, $C_R = 0.49$ ($+$), $C_L = 0.52$, $C_R = 0.48$ ($\Box$), $C_L = 0.53$, $C_R = 0.47$ ($\times$), $C_L = 0.54$, $C_R = 0.46$ ($\triangle$), $C_L = 0.55$, $C_R = 0.45$ ($*$), $C_L = 0.56$, $C_R = 0.44$ ($\diamond$), $C_L = 0.57$, $C_R = 0.43$ ($+$) and $C_L = 0.58$, $C_R = 0.42$ ($\Box$); (c) as in (a) for $C_L = C_R = 0.5$ on a basic block with 22,350 sites ($\diamond$) and a basic block with 10,940 sites ($+$) (indistinguishable).

Fig. 14. (a) Diffusion coefficient $D$ as a function of the time $t$ on a $\log_{10}$-$\log_{10}$ scale for the flipping rotator model on a Fibonacci quasi-lattice for $C_L = C_R = 0.5$ ($\diamond$), $C_L = 0.8$, $C_R = 0.2$ ($+$) and $C_L = 1.0$, $C_R = 0$ ($\Box$); (b) corresponding $\hat{P}(r,t)$ as a function of $r$ for $C_L = C_R = 0.5$ at $t = 2^{15}$ ($\diamond$), $t = 2^{17}$ ($+$) and $t = 2^{19}$ ($\Box$), respectively.

Fig. 15. (a) An example of a closed orbits for the flipping rotator model on the Fibonacci quasi-lattice with period of 1129 time steps on 274 lattice sites with two "reflectors": A and B; (b) an example of a closed orbit with period of 5997 time steps on 348 lattice sites without "reflectors". Up-triangles stand for the right rotators and down-triangles for the left rotators.

Fig. 16. (a) Fraction of closed orbits for the flipping rotator model on the triangular lattice as a function of $\log_2 t$ for different concentrations of rotators for $C_L = C_R$; from top to bottom: $C = 0.85$, 0.8, 0.7, 0.6, 0.5, 0.4, 0.3, 0.2, 0.1 (the last two are too small to see), respectively; (b) $\hat{P}(r,t)$ as a function of $r$ for the flipping rotator model (circles) and flipping mirror model (diamonds) on the triangular lattice for $C = 0.85$ and $C_L = C_R$ at $t = 2^{13}$. Smooth curves (indistinguishable) represent the corresponding Gaussian distribution (note the few points representing closed orbits near the origin).

# Table Captions

Table I: Fixed Scatterers.

Table II: Flipping Scatterers.

**TABLE I - FIXED SCATTERERS**

| Lattice | Rotator | Mirror |
|---|---|---|
| square | $0 < C \leq 1 \rightarrow$ Class IV<br>except on two critical lines[1] $\longrightarrow$ Class II | $0 < C < 1 \rightarrow$ Super-diffusion<br>$C = 1 \rightarrow$ Class II. |
| triangular | $0 < C \leq 1 \rightarrow$ Class IV<br>except on one critical line<br>$C_L = C_R$[1] $\longrightarrow$ Class II | $0 < C \leq 1 \rightarrow$ Class IV<br>except on one critical line<br>$C_L = C_R$[1] $\longrightarrow$ Class II |
| honeycomb | $C = 1 \rightarrow$ Class IV<br>except for two critical points $\longrightarrow$ Class II | $C = 1 \rightarrow$ Class IV |
| quasi | $C = 1 \rightarrow$ Class IV | |

**TABLE II - FLIPPING SCATTERERS**

| Lattice | Rotator | Mirror |
|---|---|---|
| square | $0 < C < 1 \rightarrow$ Class IV | $0 < C \leq 1 \rightarrow$ Normal (Class I) |
| | $C = 1 \rightarrow$ Normal (Class I) | |
| triangular | $0 < C < 1 \rightarrow$ Quasi-normal | $0 < C < 1 \rightarrow$ Quasi-normal |
| | $C = 1 \rightarrow$ Propagation | $C = 1 \rightarrow$ Propagation |
| honeycomb | $C = 1 \rightarrow$ Class IV | $C = 1 \rightarrow$ Class IV |
| quasi | $C = 1 \rightarrow$ Class IV | |

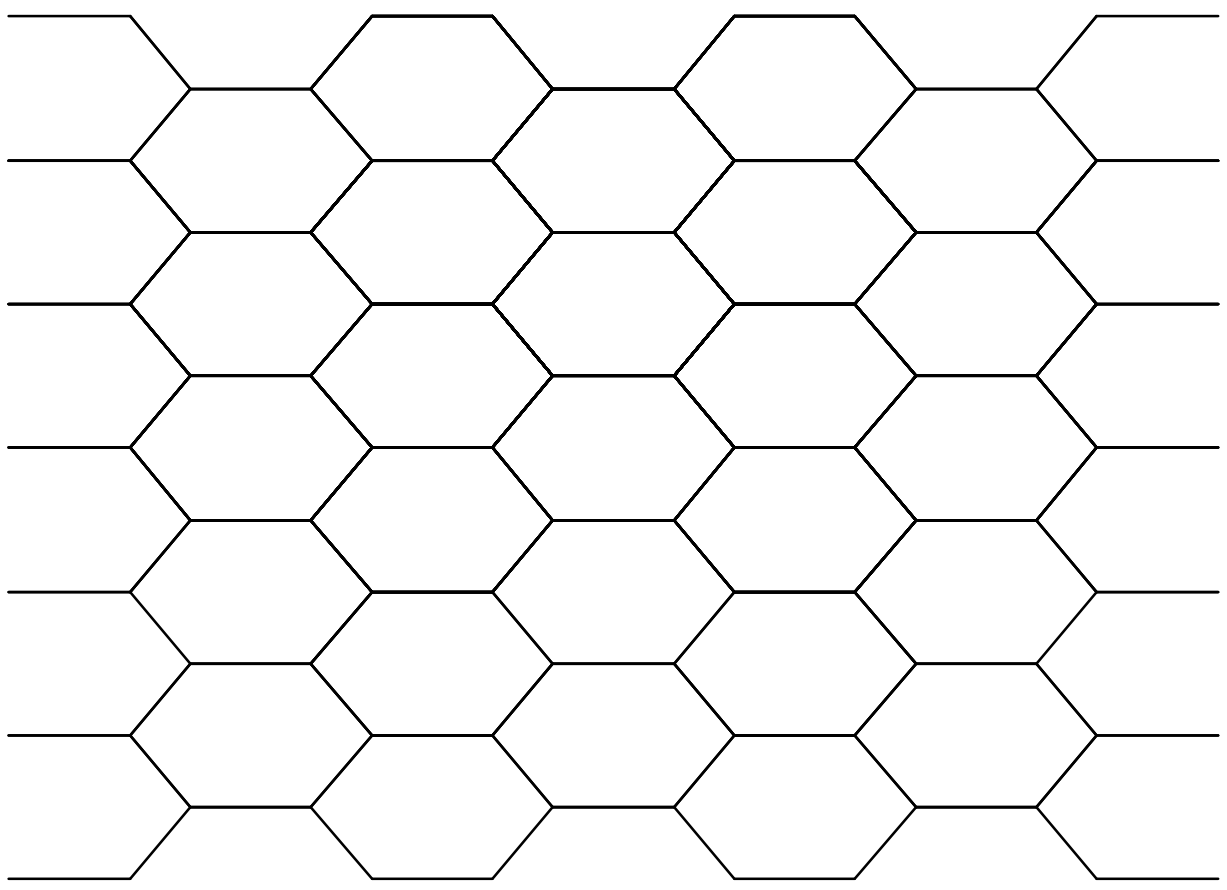

Fig. 1

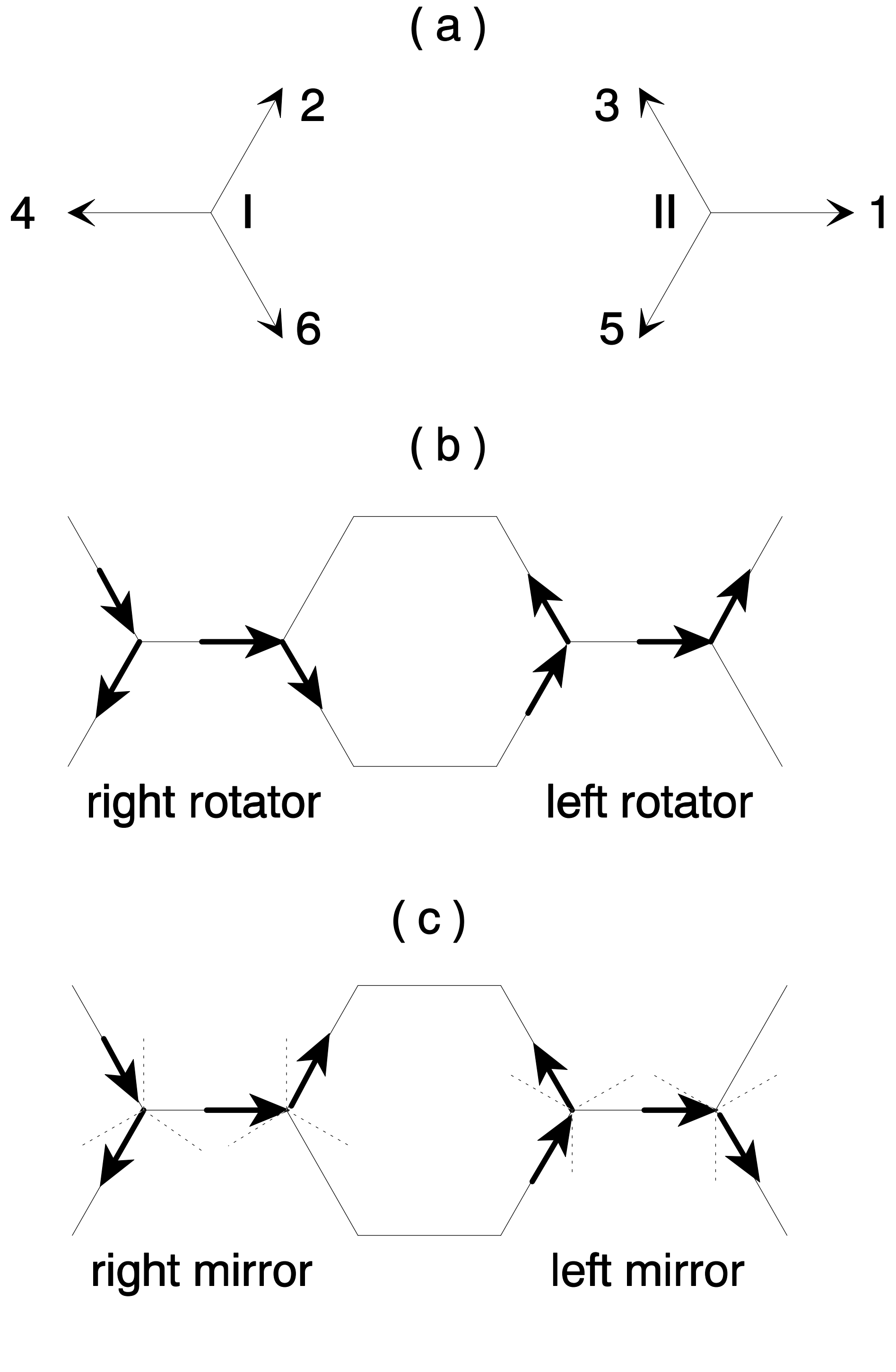

Fig. 2

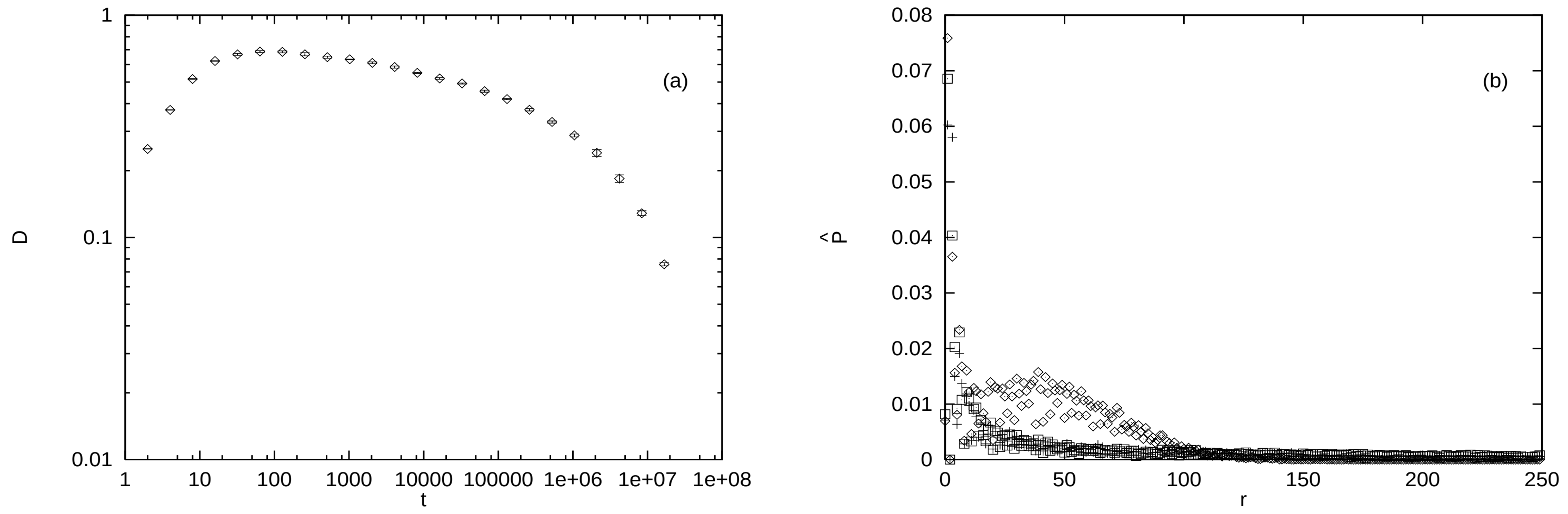

Fig. 3

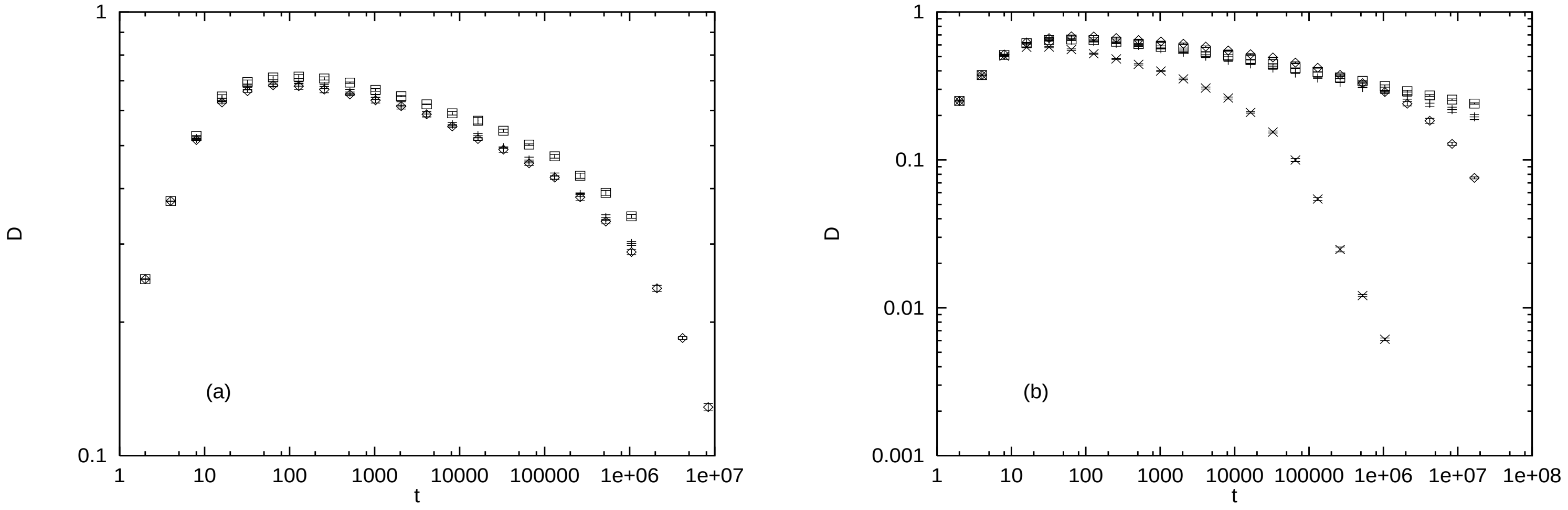

Fig. 4

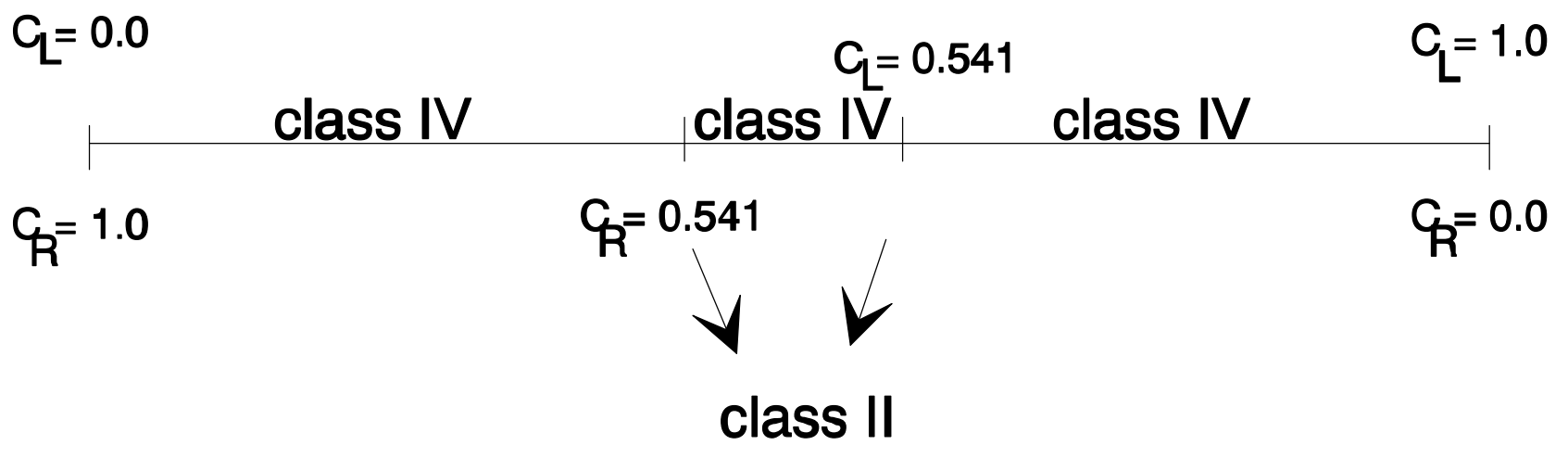

Fig. 5

.

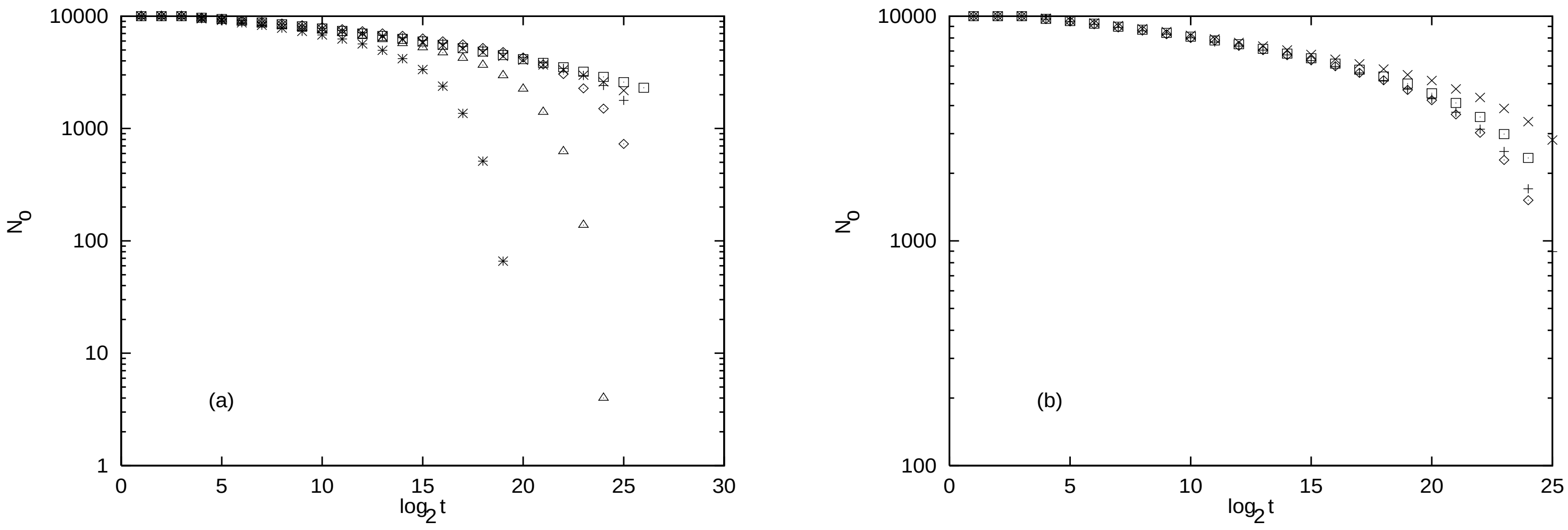

Fig. 6

.

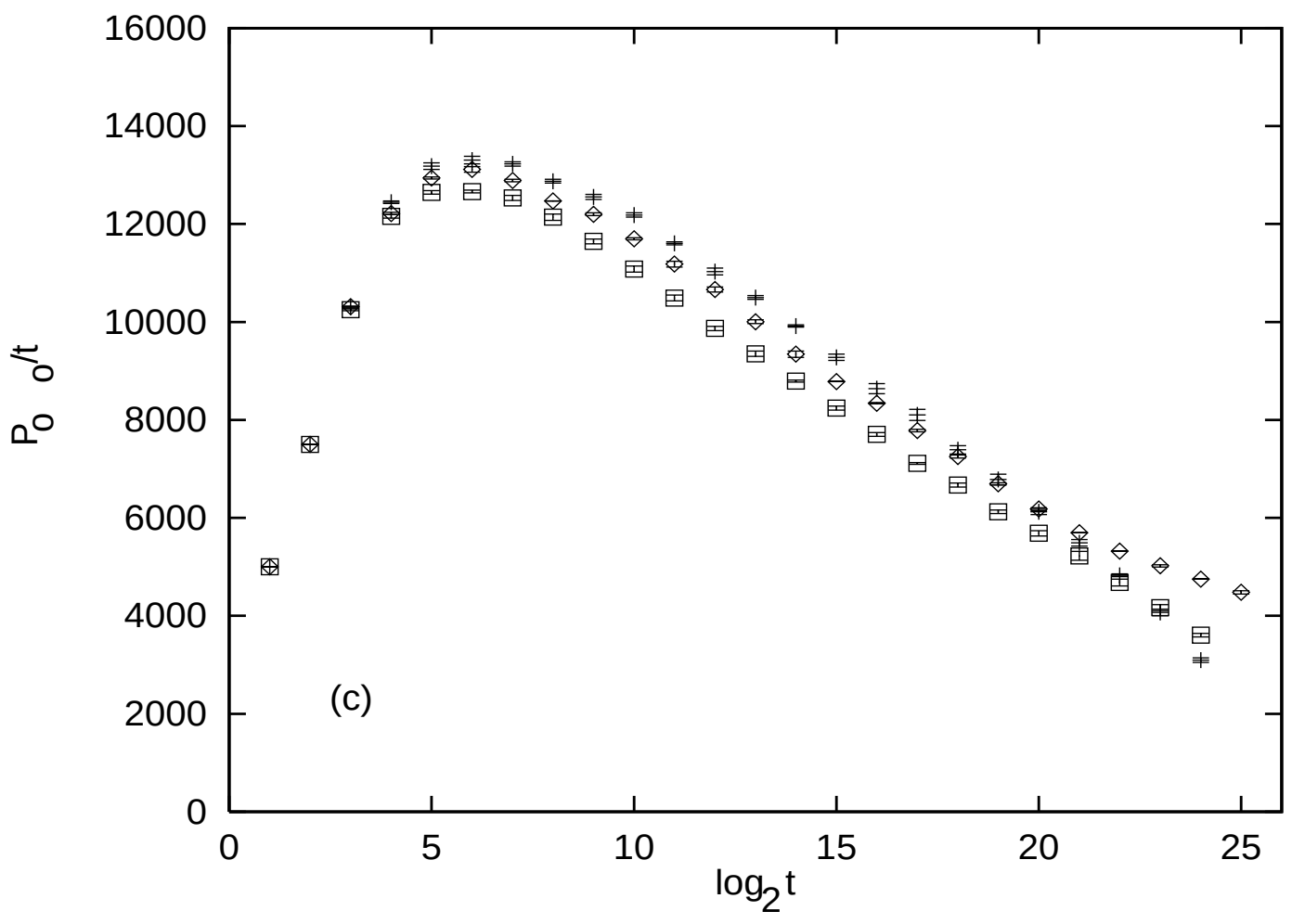

Fig. 6

.

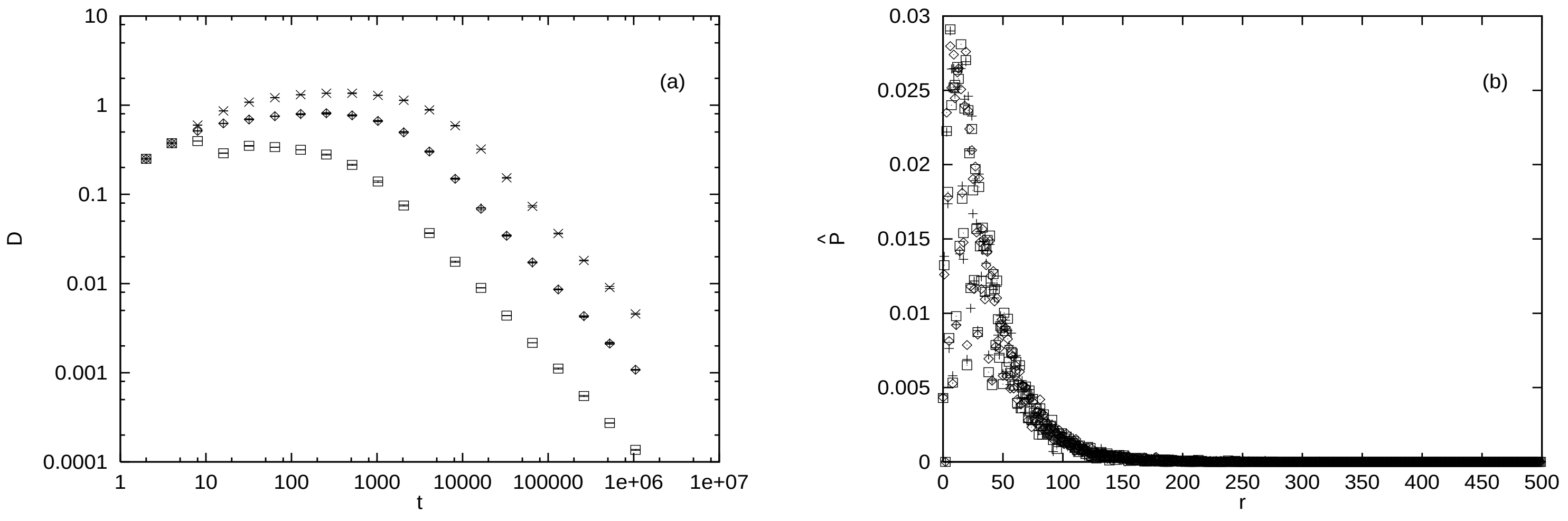

Fig. 7

.

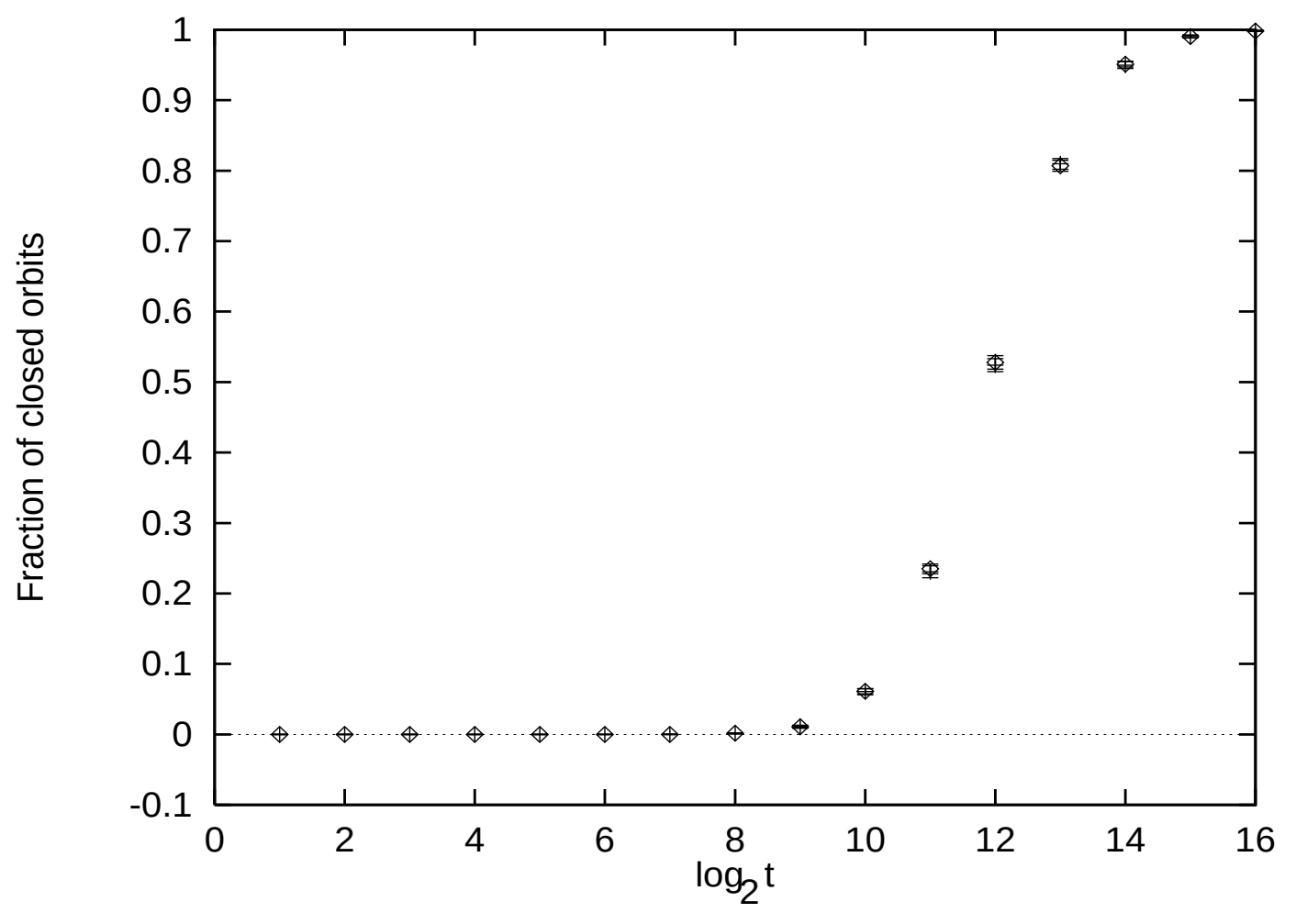

Fig. 8

.

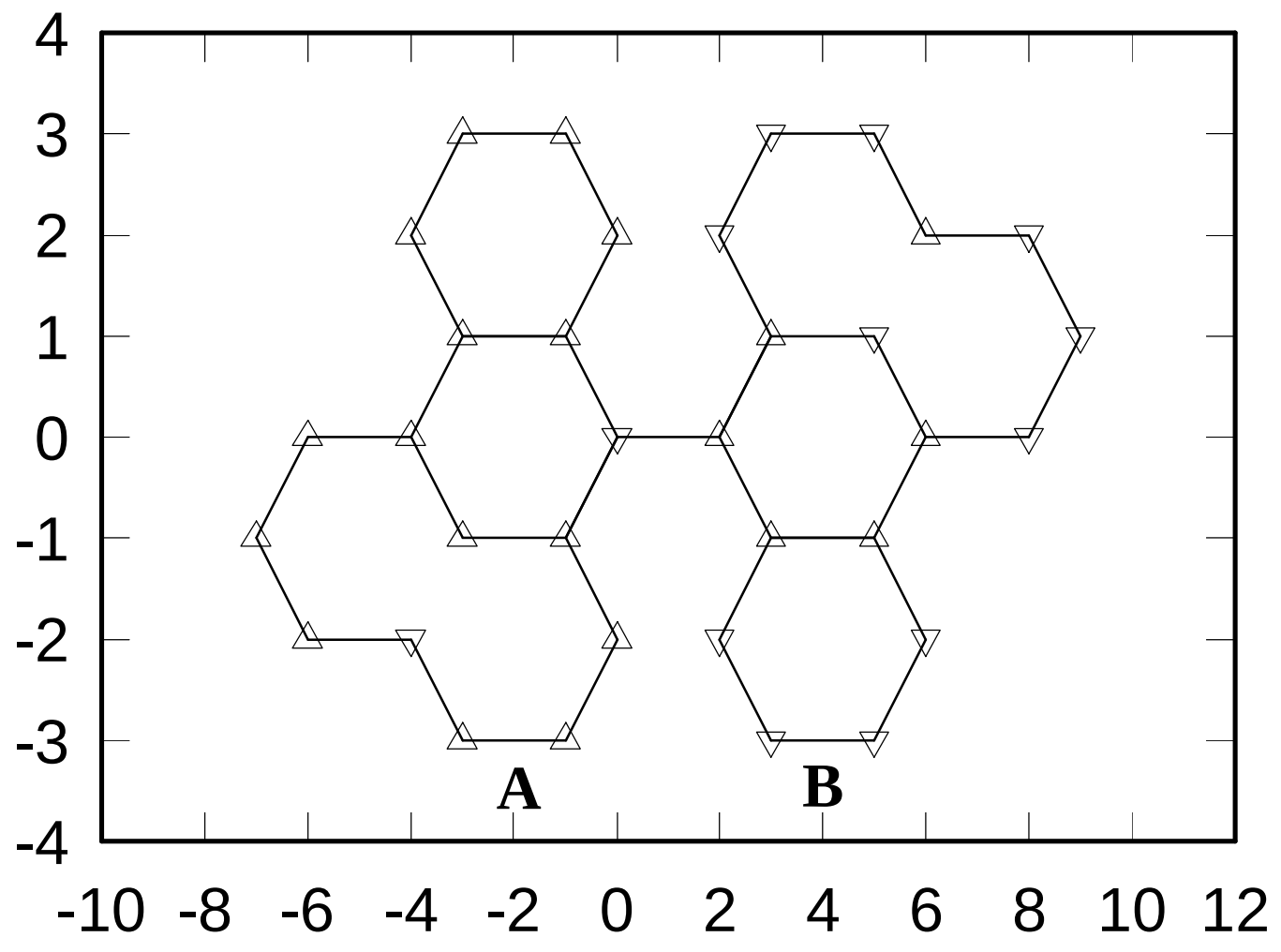

Fig. 9

**(a)**

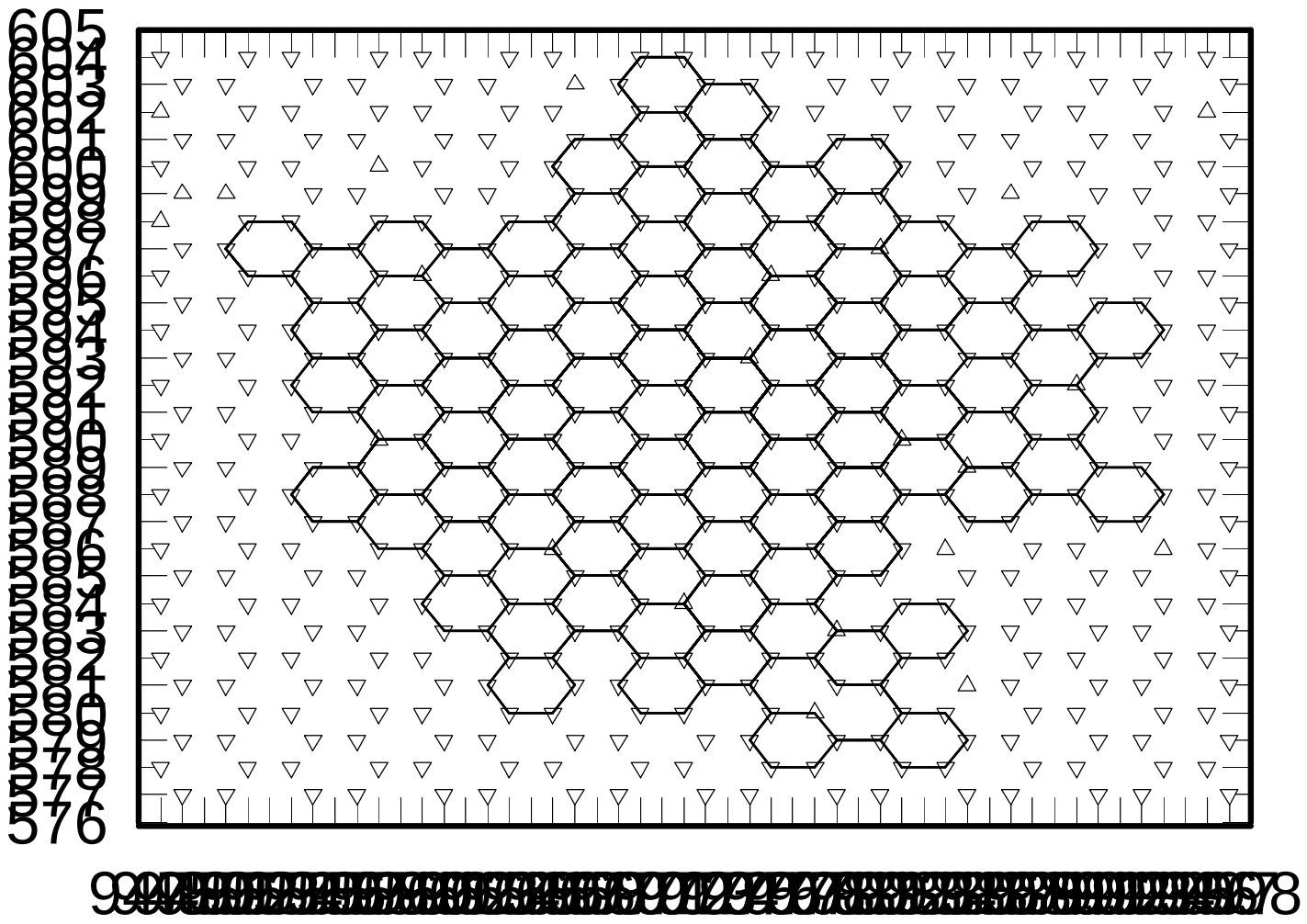

Fig. 10

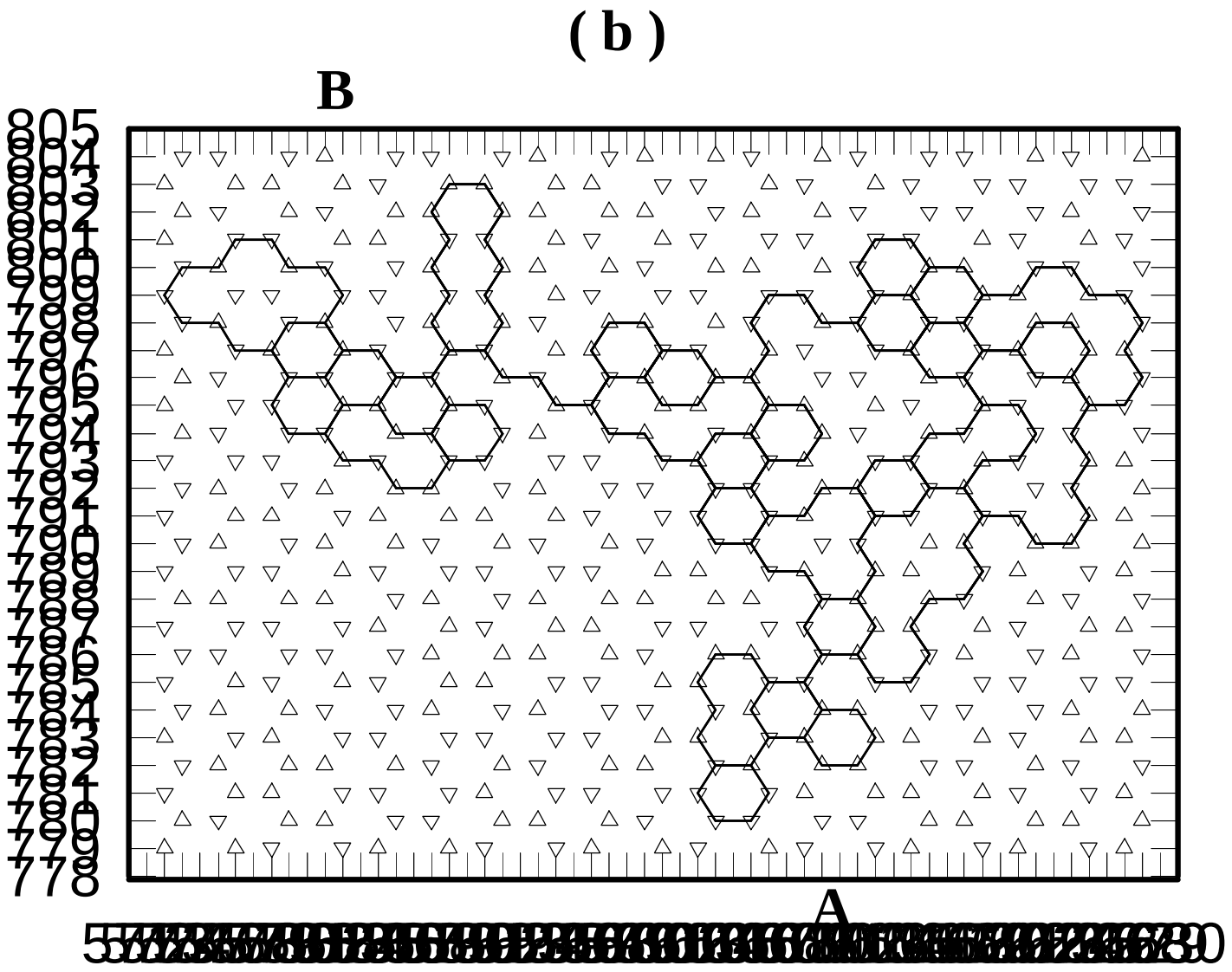

Fig. 10

( c )

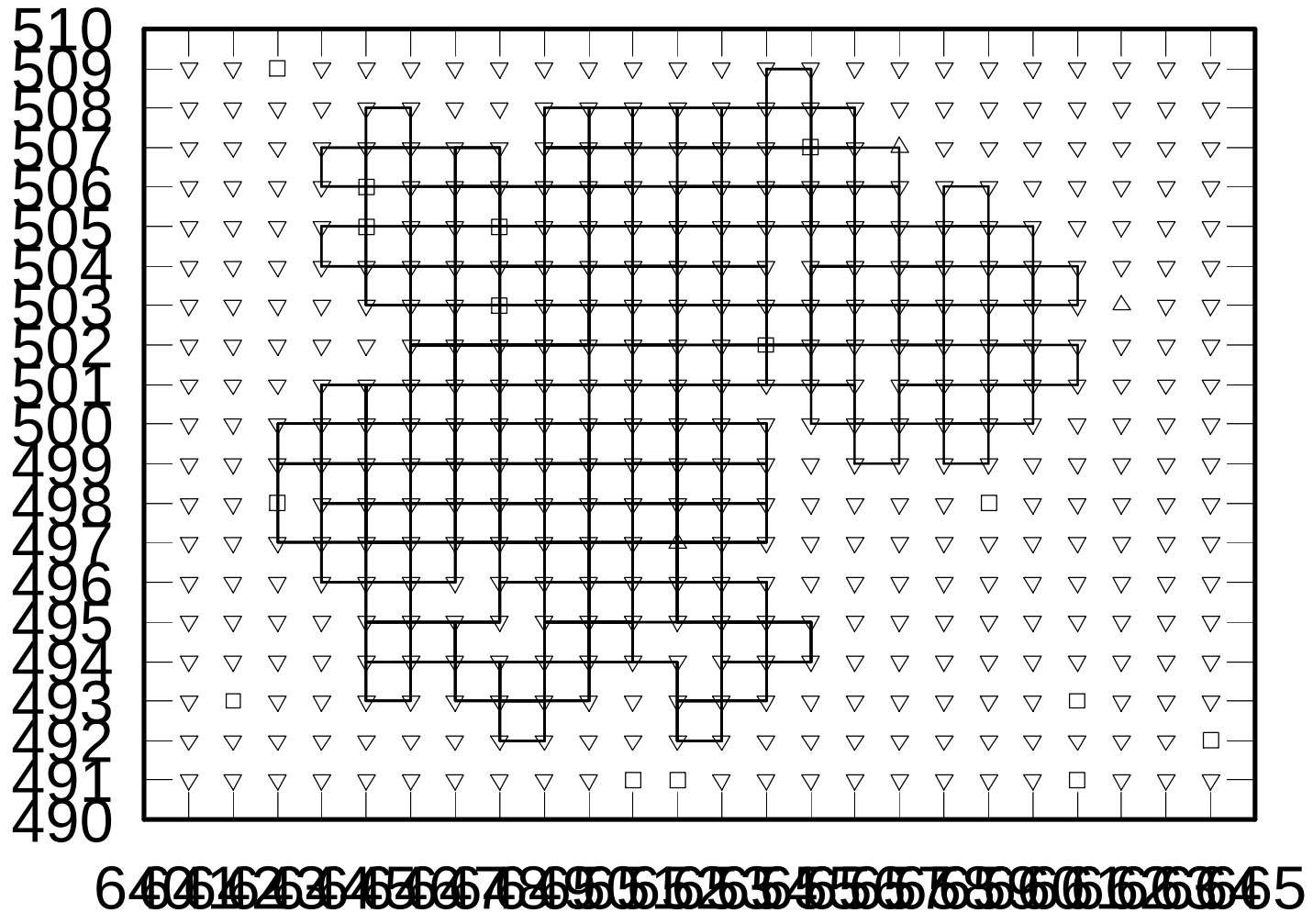

Fig. 10

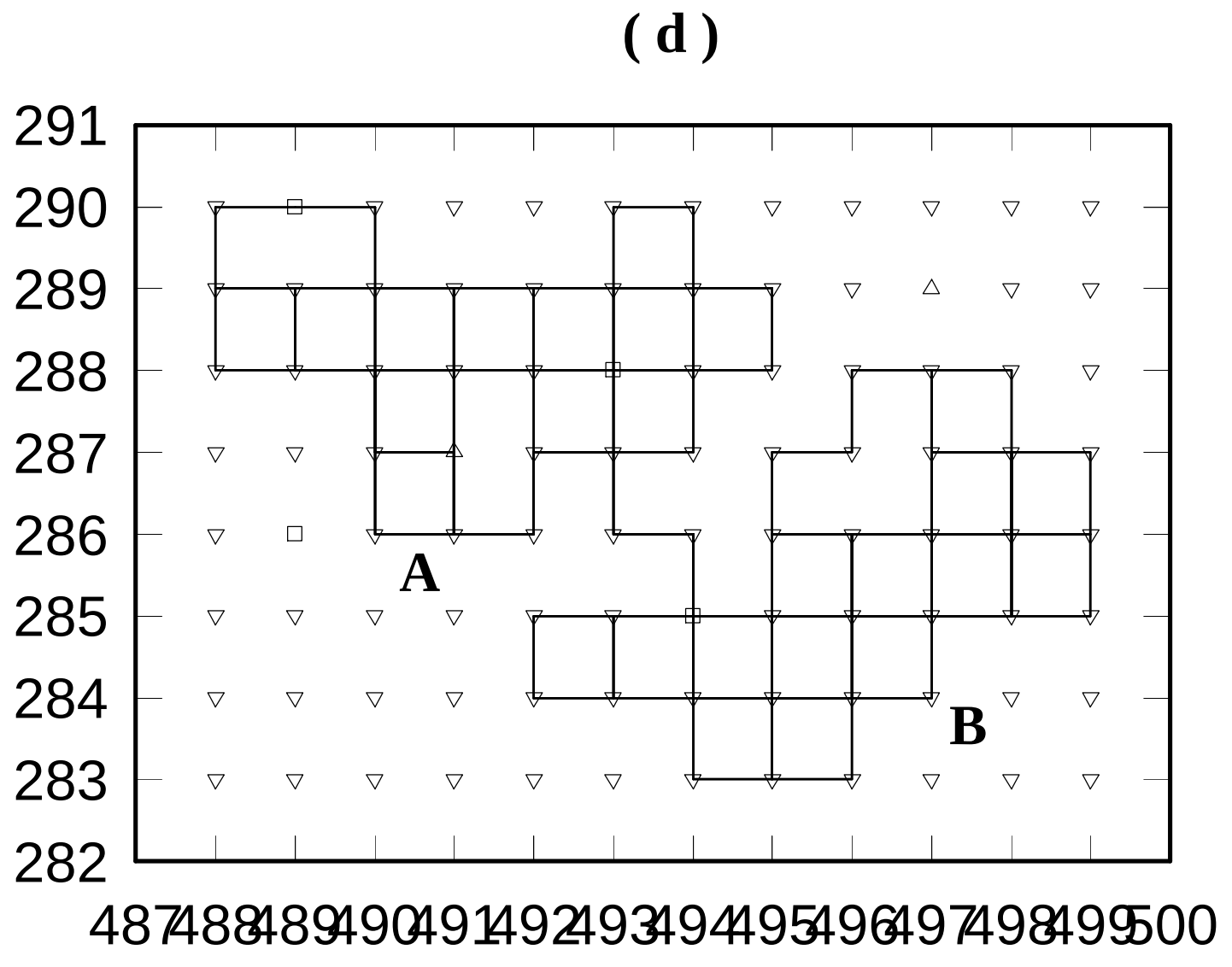

Fig. 10

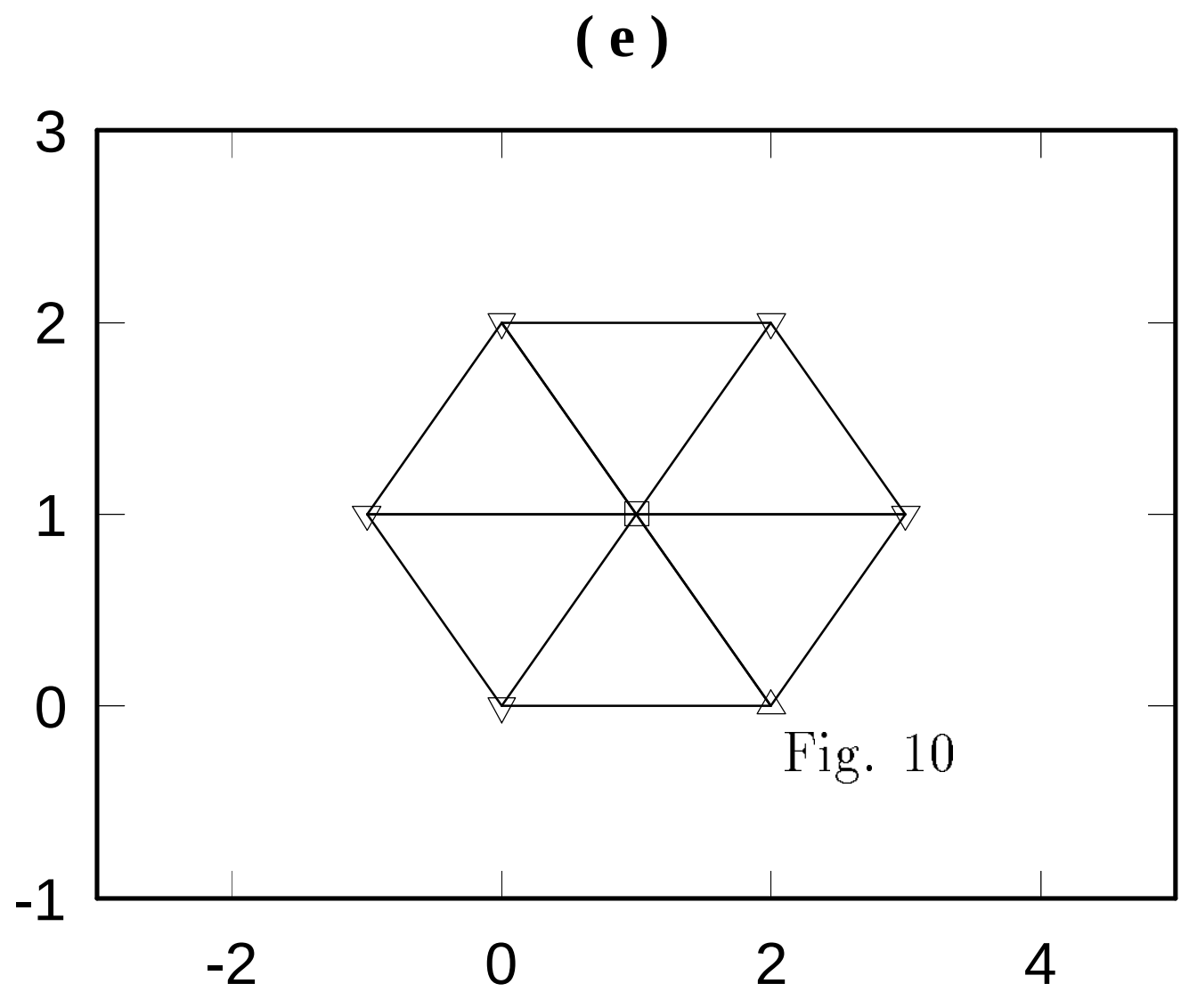

Fig. 10

Fig. 10

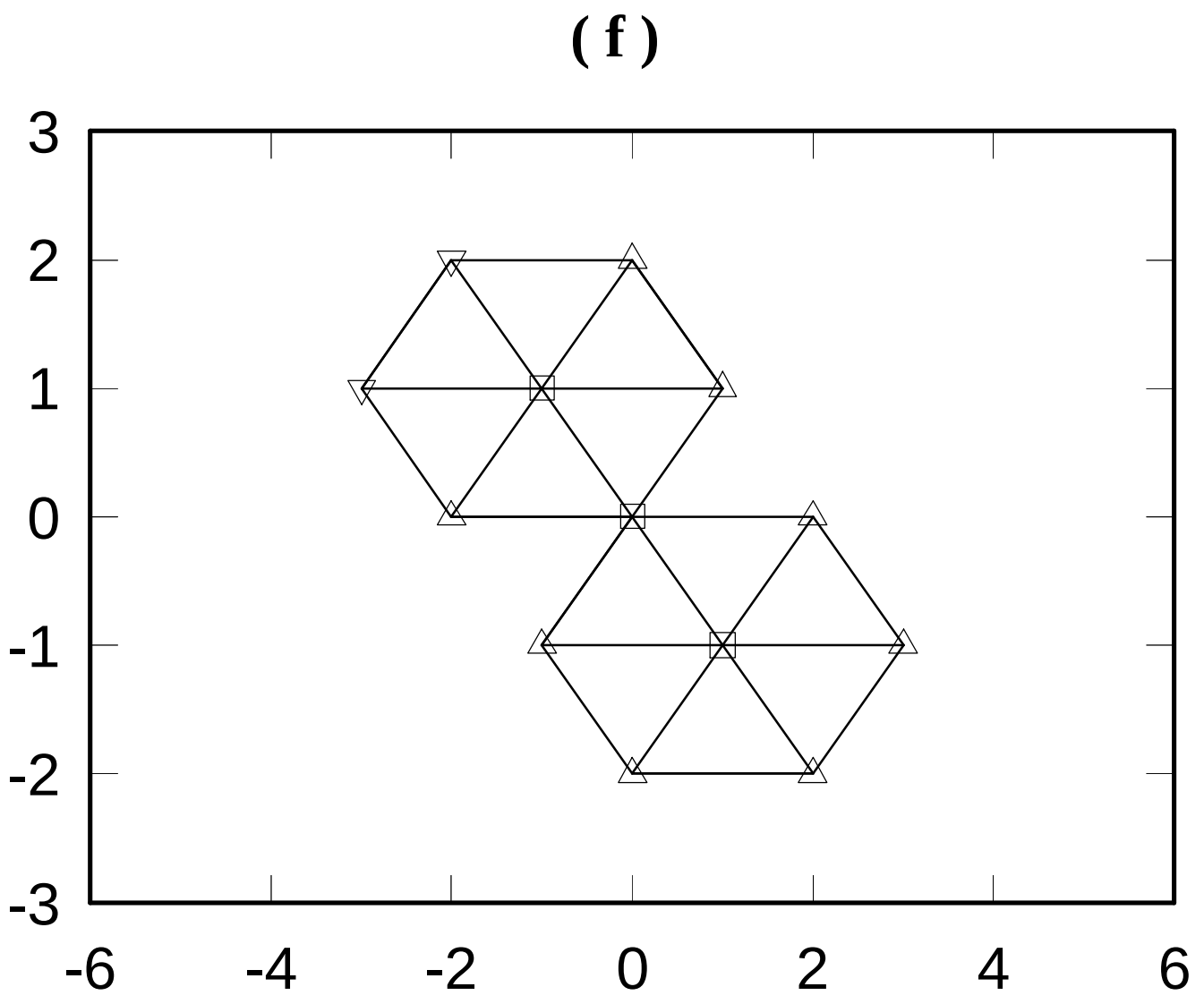

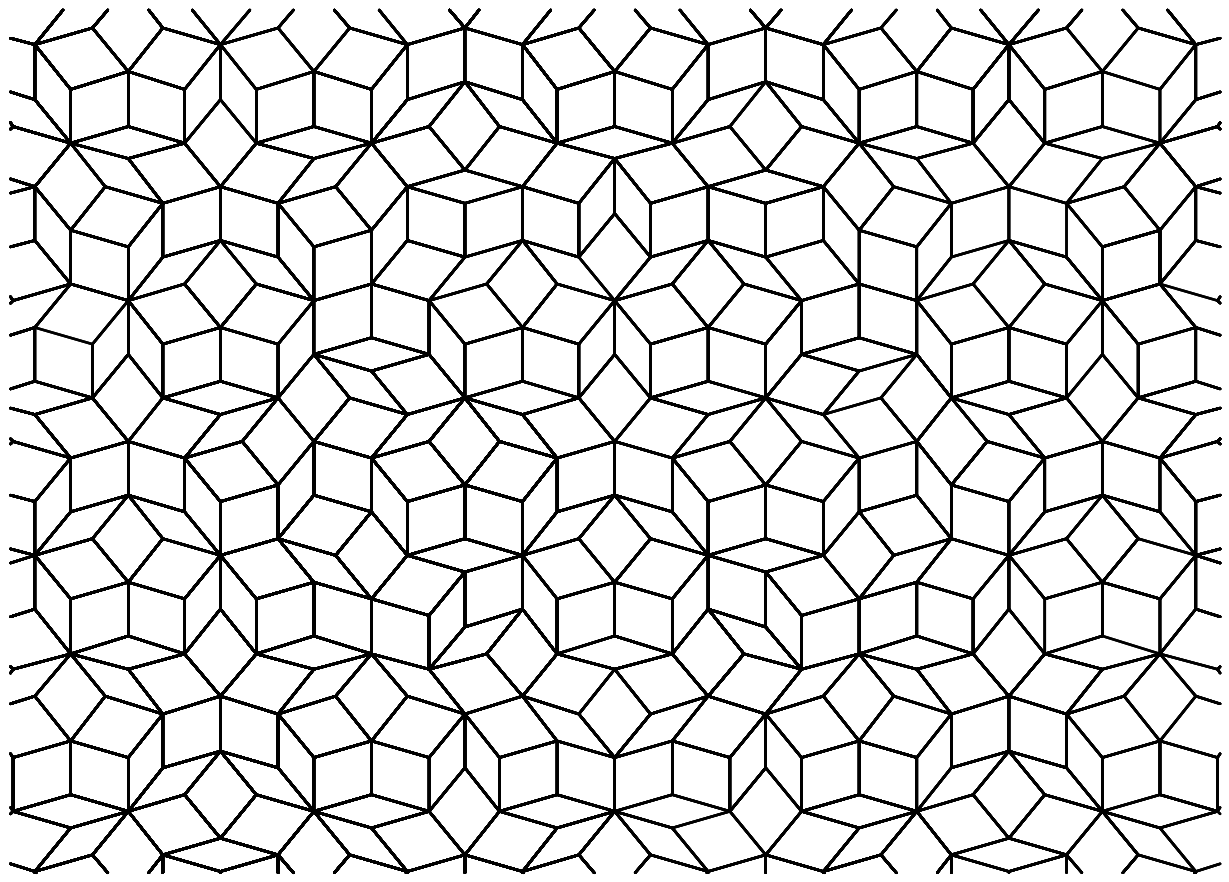

Fig. 11

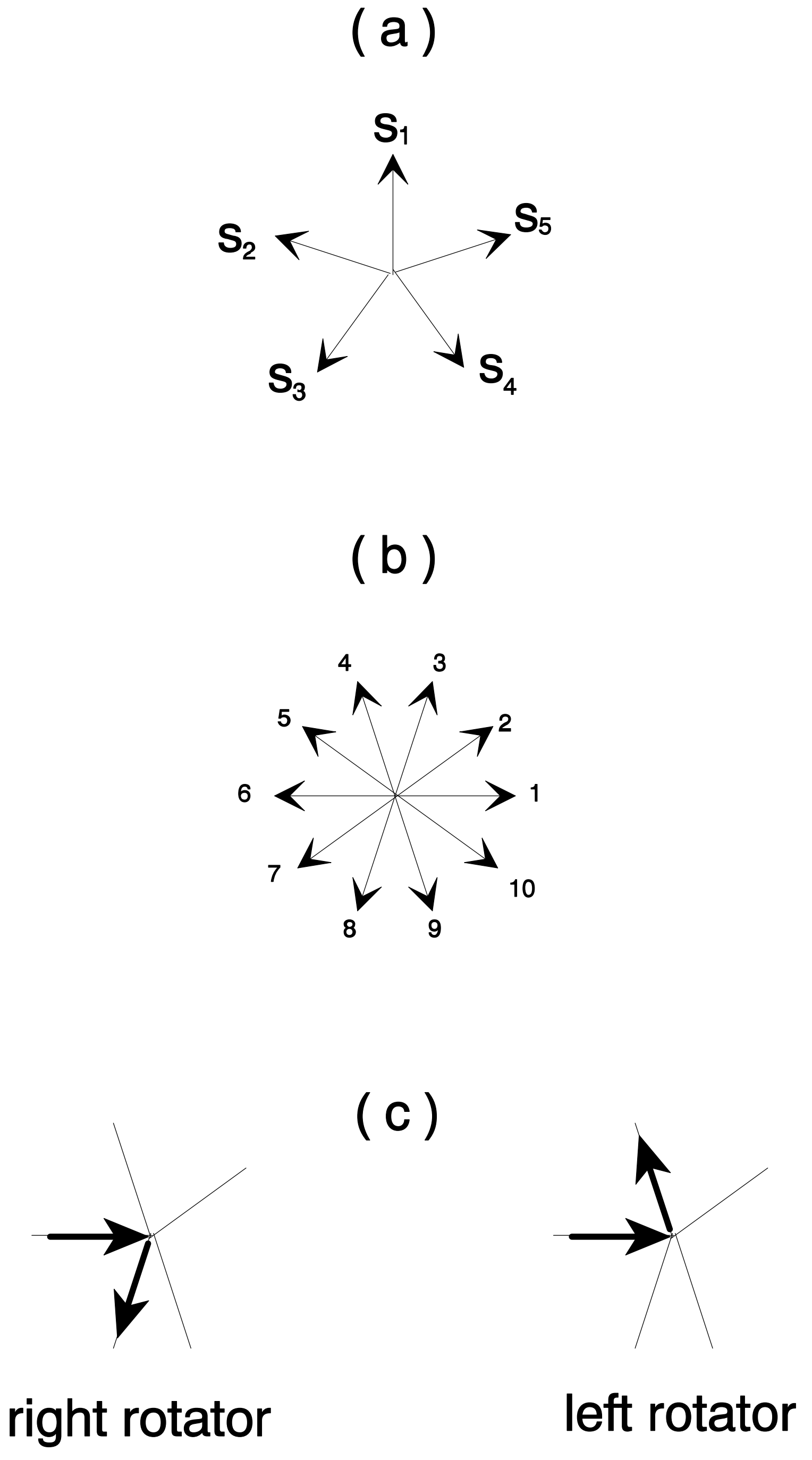

Fig. 12

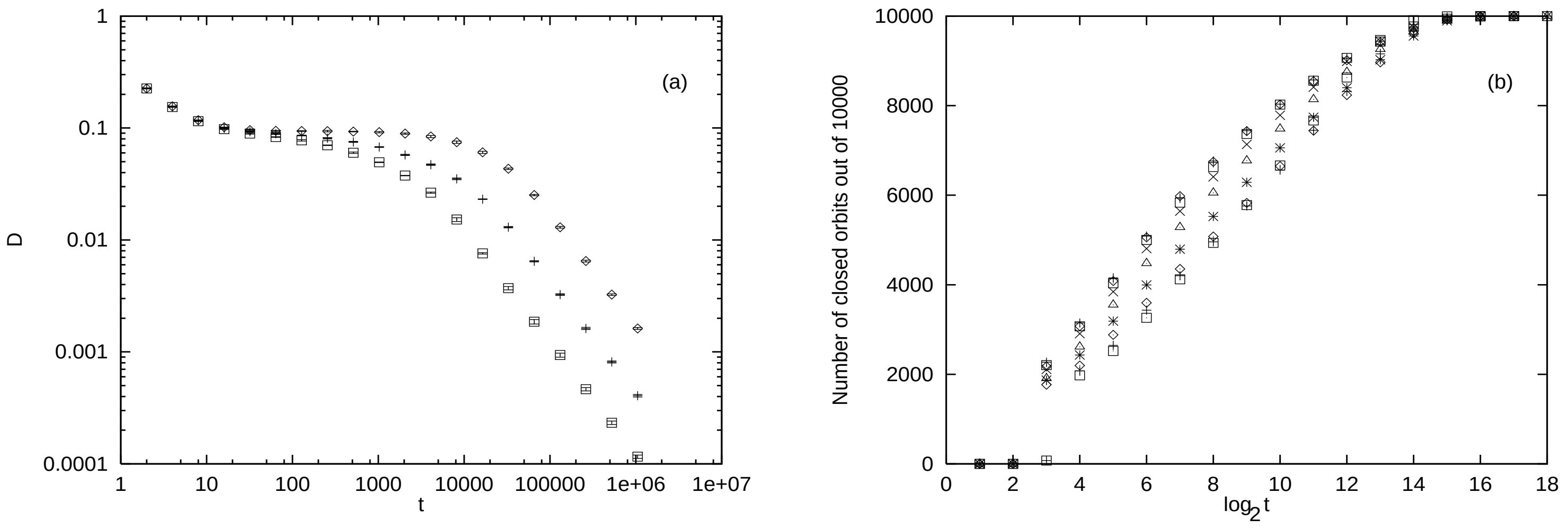

Fig. 13

.

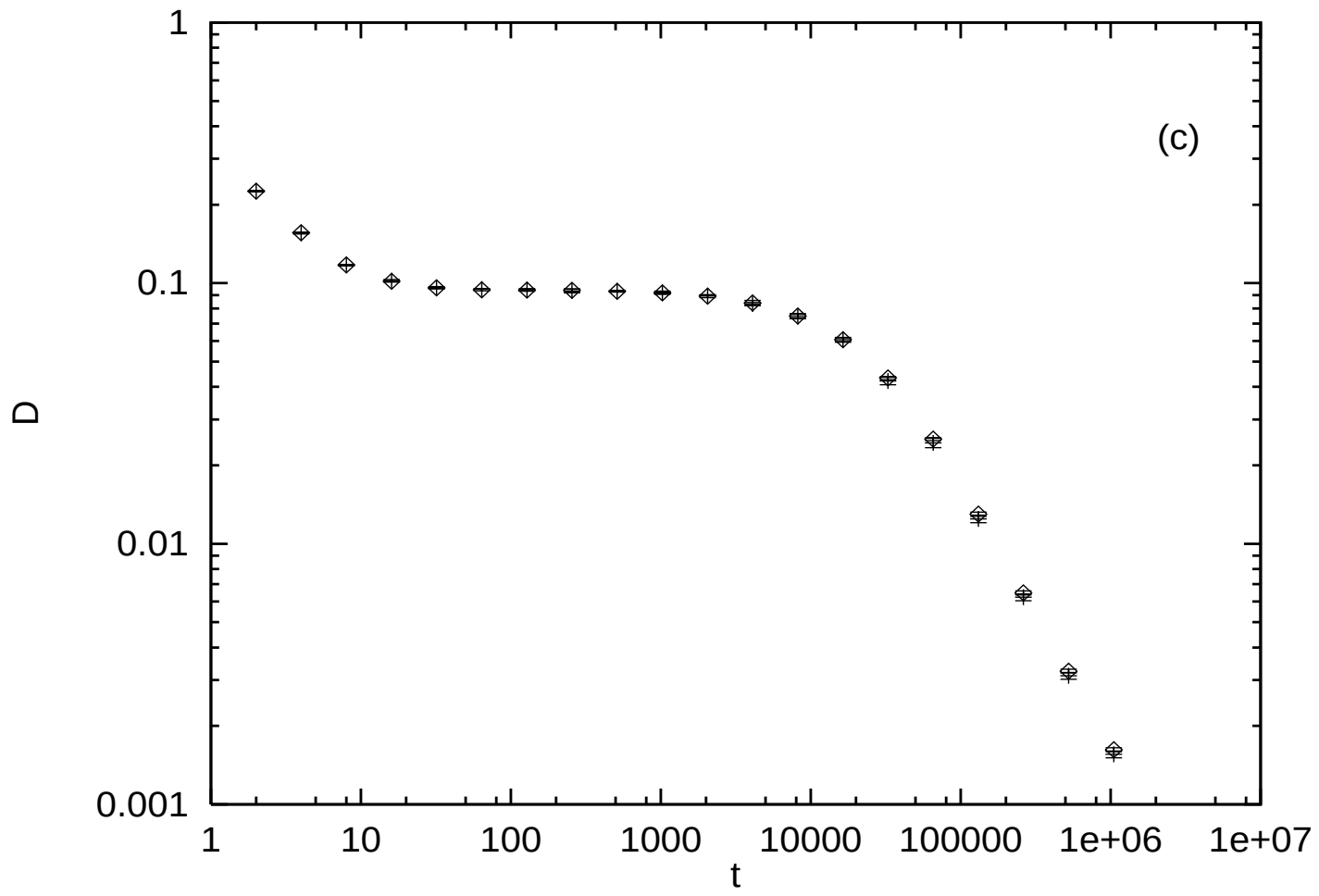

Fig. 13

.

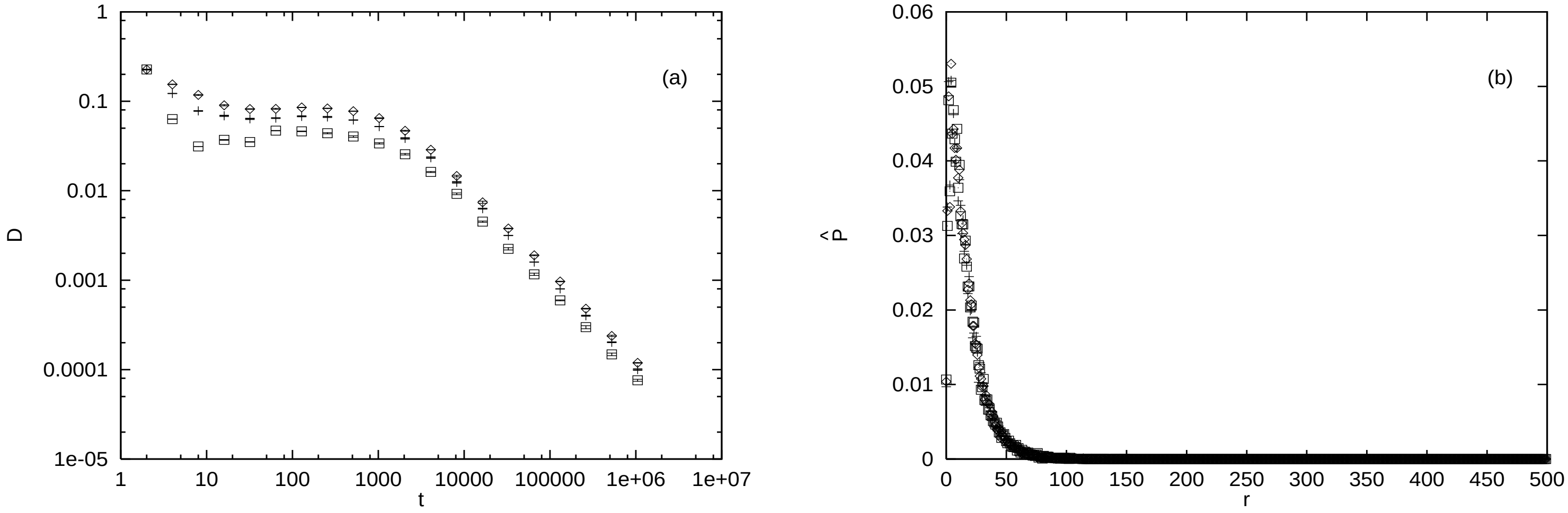

Fig. 14

**( a )**

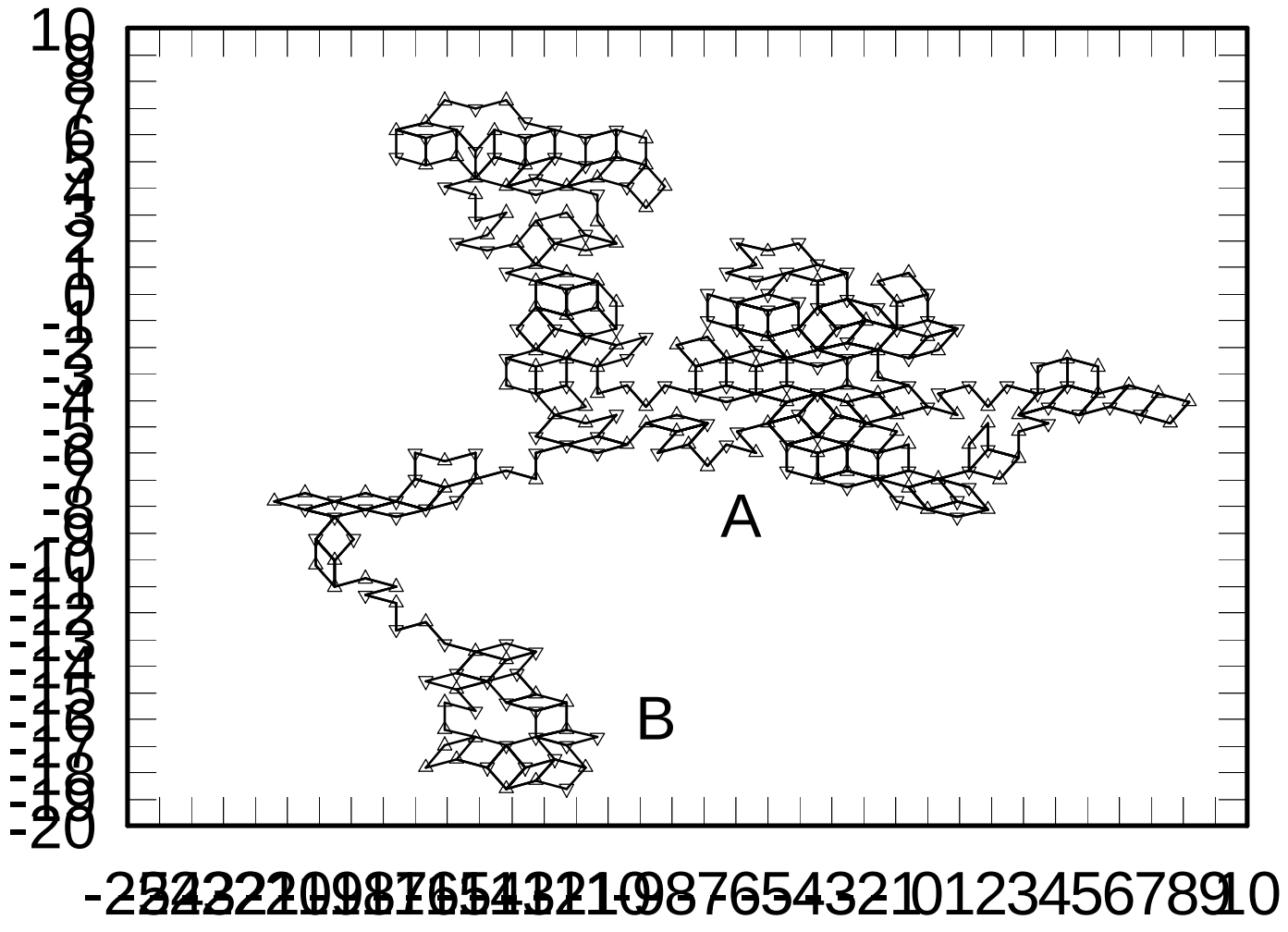

Fig. 15

( b )

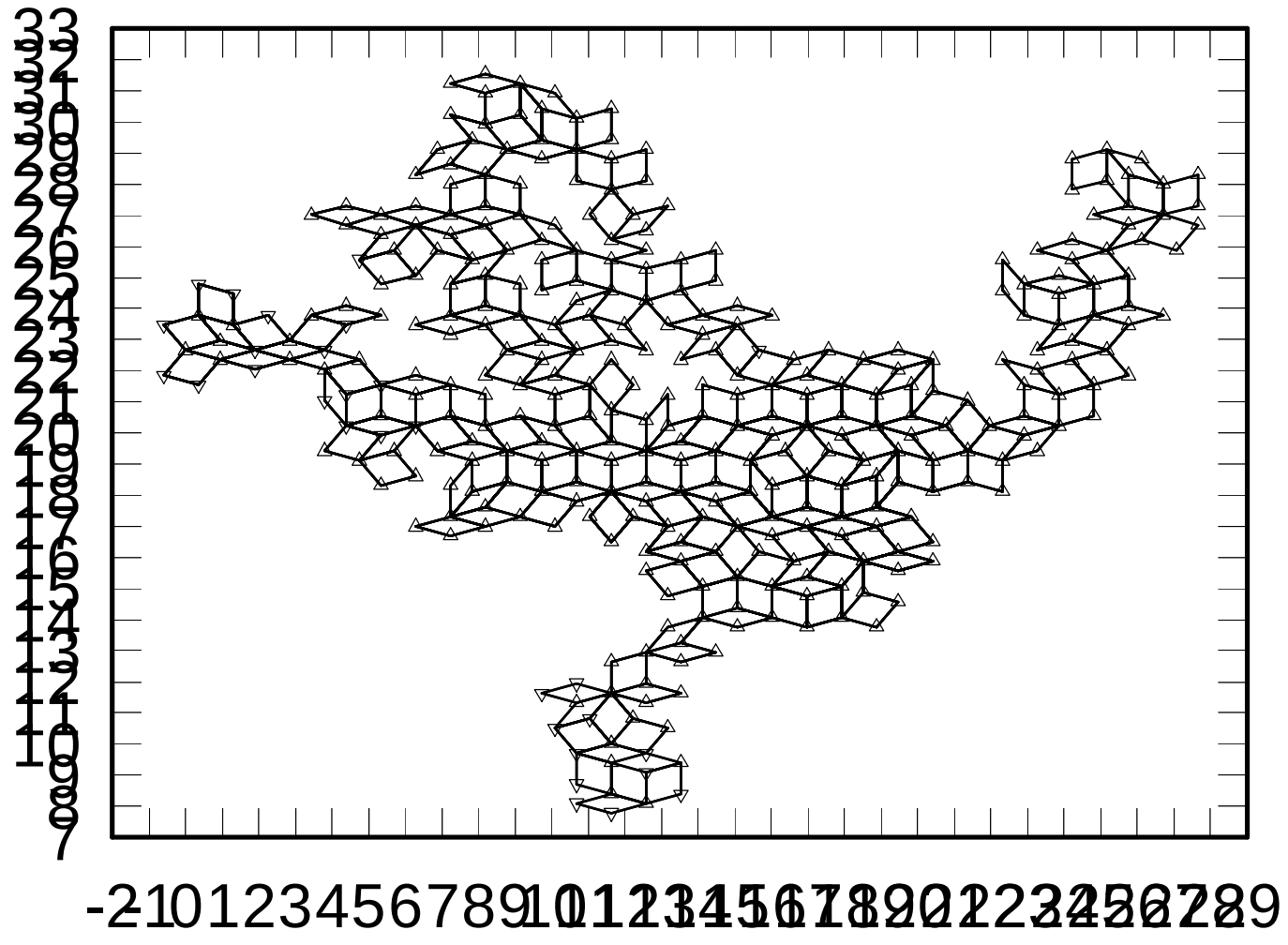

Fig. 15

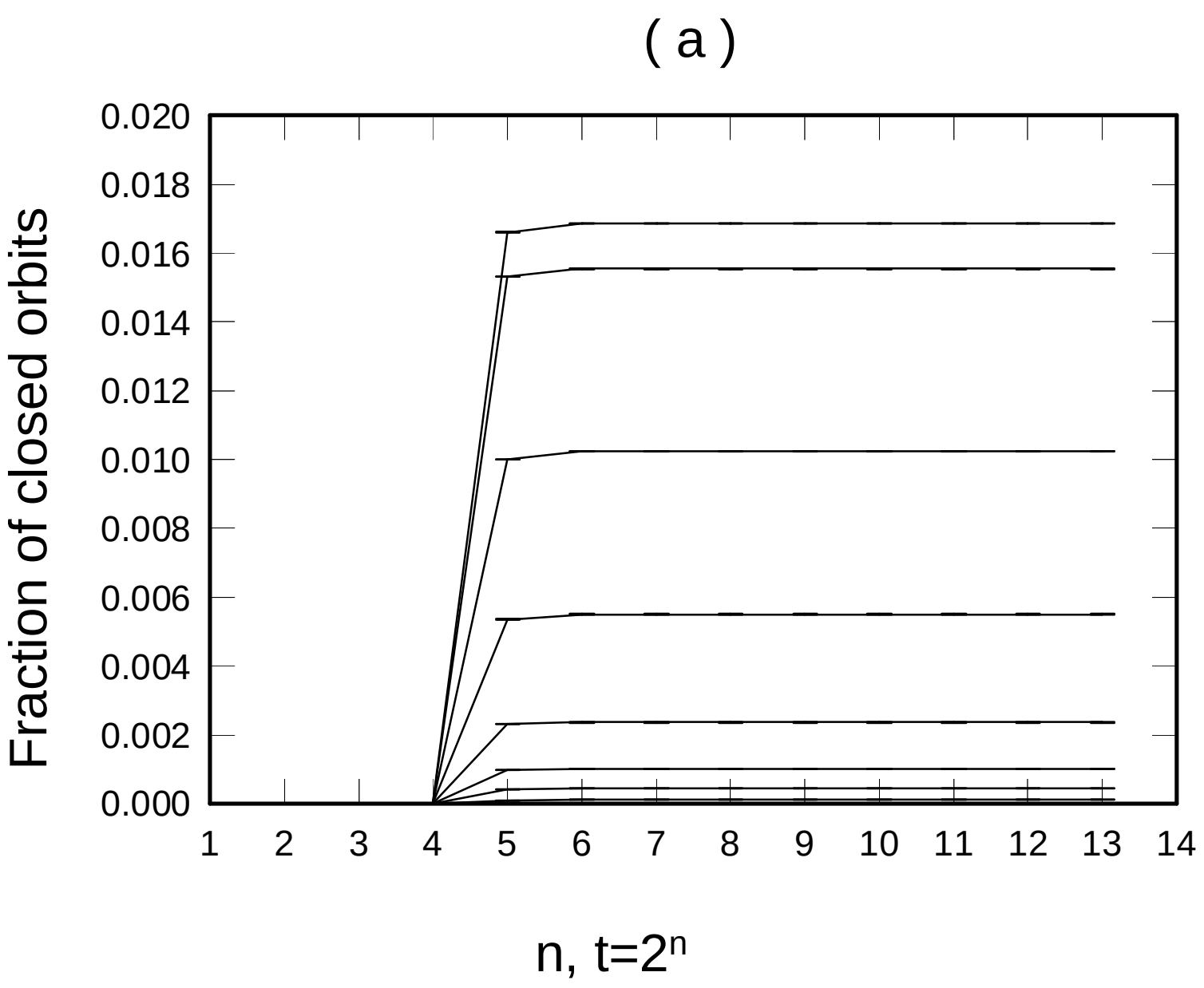
( a )
Fraction of closed orbits
0.020
0.018
0.016
0.014
0.012
0.010
0.008
0.006
0.004
0.002
0.000
1 2 3 4 5 6 7 8 9 10 11 12 13 14
n, t=2ⁿ

Fig. 17

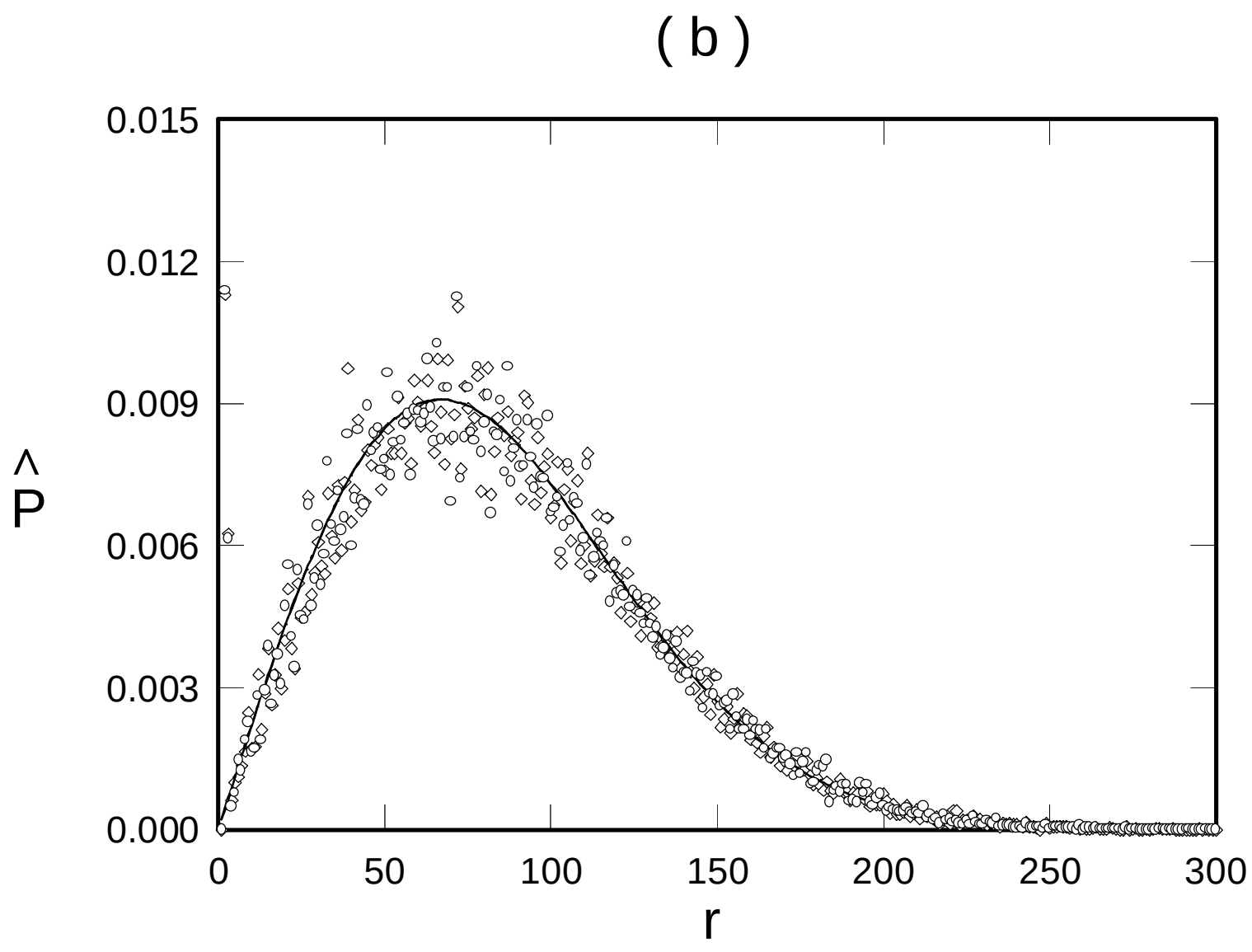

Fig. 17